\documentclass[mathleft,fleqn,%
]{an}
\usepackage{graphicx}
\usepackage{amsmath}
\usepackage[varg]{txfonts}
\begin{document}

% The following seven commands are intended for editorial usage and
% should be ignored by the author(s).
\Pagespan{1}{}% Document's page range. 
% If second parameter is left empty, the last page is computed
% automatically.
\Yearpublication{2014}%
\Yearsubmission{2014}%
\Month{0}%   
\Volume{999}%  
\Issue{0}% 
\DOI{asna.201400000}% 

\title{Phase coherence and phase jumps in the Schwabe cycle}

\author{F. Stefani\inst{1}\fnmsep\thanks{Corresponding author:
        {F.Stefani@hzdr.de}}
% Example for footnote, note the usage of the \texttt{fnmsep} command
% as separator between institute number and footnote mark}
\and J. Beer\inst{2}
\and  A. Giesecke\inst{1}
\and T. Gloaguen\inst{3}
\and  M. Seilmayer\inst{1}
\and  R. Stepanov\inst{4}
\and  T. Weier\inst{1}}
\titlerunning{Phase coherence and phase jumps in the Schwabe cycle}
\authorrunning{F. Stefani et al.}
\institute{
Helmholtz-Zentrum Dresden-Rossendorf, Bautzner Landstr. 400, 01328 Dresden, Germany
\and
Eawag, \"Uberlandstrasse 133, 8600 D\"ubendorf, Switzerland
\and
Martin-Andersen-Nex\"o-Gymnasium, Haydnstra\ss e 49, 
D-01309 Dresden, Germany
\and 
Institute of Continuous Media Mechanics, 1 Acad. Korolyov 
str., 614013 Perm, Russia}

\received{XXXX}
\accepted{XXXX}
\publonline{XXXX}


\abstract{Guided by the working hypothesis that the Schwabe cycle
of solar activity is synchronized by the 11.07 years alignment 
cycle of the tidally dominant planets Venus, Earth and Jupiter, 
we reconsider the phase diagrams 
of sediment accumulation rates 
in lake Holzmaar, and of methanesulfonate (MSA) data in
the Greenland ice core GISP2,  
which are available for the period 10000-9000 cal. BP. Since 
some half-cycle phase jumps appearing in  the 
output signals are, very likely, artifacts of applying
a biologically substantiated 
transfer function, the underlying 
solar input signal with a dominant 
11.04 years periodicity
can be considered as 
mainly phase-coherent over the 1000 years period in the 
early Holocene. 
For more recent times, we show that the re-introduction
of a hypothesized ``lost cycle''
at the beginning of the Dalton minimum
would lead to a real phase jump. Similarly, by analyzing various
series of $^{14}$C and $^{10}$Be data and comparing them with 
Schove's historical cycle maxima, we support the existence of another 
 ``lost cycle'' around 1565, also connected 
with a real phase jump. 
Viewed synoptically, our results lend greater plausibility to the 
starting hypothesis of a tidally synchronized solar cycle, which 
at times can undergo phase jumps, although the competing 
explanation in terms of a non-linear solar dynamo 
with increased coherence cannot be completely ruled out.
  }

\maketitle

\section{Introduction}
In hindsight, it is a mystery that the important paper of 
Vos et al. (2004) has been so widely overlooked in the solar physics 
community. 
Based on a careful selection of the  
10000-9000 cal. BP segment of varved sediment accumulation 
data from lake 
Holzmaar (Vos et al. 1997), and employing a nonlinear transfer function reflecting the
dependence of the biological productivity of autochthonous algae  
on temperature and/or 
solar radiation (in particular its UV component), the authors 
revealed strong evidence for a Schwabe cycle with a dominant 11.04 years 
period which was basically 
phase coherent over the considered 1000-year interval. 
The same 
periodicity and phase coherence was also detected in GRISP2  ice-core
data, when analyzing the methanesulfonate (MSA) concentration as a marker 
of algal productivity in the North Atlantic. Some 
phase jumps by half a cycle (i.e. 5.5 years), showing up 
in both time series, appeared as artifacts of applying
the transfer function, so that ultimately the input data of the 
underlying solar cycle 
was interpreted  
as phase coherent over 1000 years. 

While such a long phase coherence, observed in 
two rather unrelated proxy data for the Schwabe cycle, is 
most remarkably in itself, 
one might be even more puzzled by the almost complete equivalence of 
the cycle period of 11.04 years with the corresponding 
11.07-year period as derived for  
the last centuries (Stefani et al. 2019, 2020). Ironically, 
the latter value for the more recent times 
is more disputable than the value for the early Holocene, 
given that the systematic  
observation of sunspots goes back 
only to the times of Scheiner and Galileo 
(Arlt and Vaquero 2020), with
grave uncertainties for the time of the Maunder minimum 
(1645-1715).

In an unparalleled effort of analyzing historical {\it aurora borealis} 
observations, Schove (1955, 1979, 1983) had tried to extend 
the widely accepted series of cycle minima and maxima to a much 
longer period of nearly two and a half millennia. 
The resulting time series prompted different reactions, 
ranging from appreciations of their high quality (Jelbring 1995) 
to less enthusiastic qualifications as being ``archaic'' 
(Usoskin 2017). With all due caveats about their reliableness,
we recently analyzed Schove's cycle maxima 
data down to A.D.~240 (where a first data gap appears),
and found a clearly dominant and largely phase coherent 
cycle with a 11.07-years periodicity 
(Stefani et al. 2020), modulated 
by two longer cycles of the Suess-de Vries type (around 200 years) 
and the Gleissberg type (around 90 years).
Moreover, analyzing Dicke's ratio 
$\sum_i r_i^2/\sum_i (r_i-r_{i-1})^2$ (Dicke 1978) 
between the mean square of the residuals $r_i$
(defined as the distances between the actual 
minima and the hypothetical minima of a 
perfect 11.07-year cycle) 
to the mean square of the differences $r_i-r_{i-1}$
between two consecutive 
residuals, the 
solar cycle was shown to have much closer resemblance to a clocked 
process than to a random walk process (Stefani et al. 2019). 
This gave further support for our conjecture
(Stefani et al. 2016, 2017, 2018, 2019)
that the Schwabe cycle results from synchronizing a 
rather conventional $\alpha-\Omega$-dynamo 
by means of an additional 11.07-year oscillation of the $\alpha$-effect
which, in turn, is related to the helicity 
oscillation of either a kink-type ($m=1$) Tayler 
instability  in the tachocline  region (Weber 2015) or a  
($m=1$) magneto-Rossby wave (Dikpati 2017, Zaqarashvili 2018).
Building on and corroborating earlier ideas of Hung (2007), Scafetta (2012), 
Wilson (2013), and Okhlopkov (2014, 2016), the source of 
this synchronized helicity was hypothesized 
to be the 11.07-year periodic tidal ($m=2$) forcing 
of Venus, Earth and Jupiter, which are the tidally dominant 
planets in the solar system.

While the hitherto overlooked 10000-9000 cal. BP data 
of Vos et al. (2004)
seem to lend more plausibility to this tidal synchronization 
conjecture, there remain two objections against it
which have to be considered seriously. The first question  
has to do with the possibility that 
the solar dynamo may undergo a sort of ``self-synchronization'',
with a phase-amplitude correlation resulting from the 
typical quenching of $\alpha-\Omega$-dynamos when
exposed to random fluctuations of $\alpha$. As shown by 
Hoyng (1996) in a rebuttal to Dicke's question ``Is there
a chronometer hidden deep in the sun?'' (Dicke, 1978), the typical 
frequency-growth rate correlation of an $\alpha-\Omega$ dynamo 
could 
lead to a finite, yet rather long 
(hundreds of years) memory in conjunction with a stable 
phase coherence, which 
could be misinterpreted as a clocked process. 
More detailed investigations of such an increased coherence
in a simple non-linear
$\alpha-\Omega$ dynamo models
are presently under way.

A second objection is related to the questionable 
quality of Schove's minima and maxima data. While they show  
a remarkable phase coherence over one or even two millennia 
(Stefani et al. 2019, 2020), this coherence would be 
destroyed in case that Schove had forgotten (or ``smuggled in'') 
one cycle or another. Sections 3 and 4 will deal with 
such ``lost cycles'' and the real 
phase jumps connected with them. 
However, before entering this issue
we will shortly discuss some fictitious phase jumps appearing
in the data of Vos et al. (2004).

\section{Algae in the sun}

Due to its key relevance to the synchronization
theory of the solar dynamo,  we reproduce Figure~17.7 of
Vos et al. (2004) in our Figure~1. It shows the
phase diagrams of varve thicknesses from Lake Holzmaar (a),
and of methanesulfonate (MSA) influx
measured in the Greenland ice core GISP2 (b), which 
had been produced by 
correlating the measured signal within a given 
comparison window (100 years)
with an ideal sinusoidal signal of known periodicity
and phase (more details about the creation of phase 
diagrams can be found in Appendix A).
Both phase diagrams reveal the 
existence of two distinct band structures, phase shifted
by half a cycle length (5.5 years), 
each of them lasting over intervals 
of one or a few hundred years. While, in general, the 
Holzmaar and the GISP2 data have a very
similar appearance, their respective phase jumps are shifted
systematically by 41 years due to the use 
of two different INTCAL calibration curves (Vos 2020).
Later, we will also come back to the remarkable
triangular structure which is visible in the GRISP2 data  between
9800 and 9700 cal. BP, but not in the corresponding Holzmaar data. 
This point will also be critically assessed in 
Appendix A, where we elaborate 
on  some sensitive parameter dependencies
of phase diagrams such as Figure~1.

\begin{figure}
\includegraphics[width=0.99\linewidth]{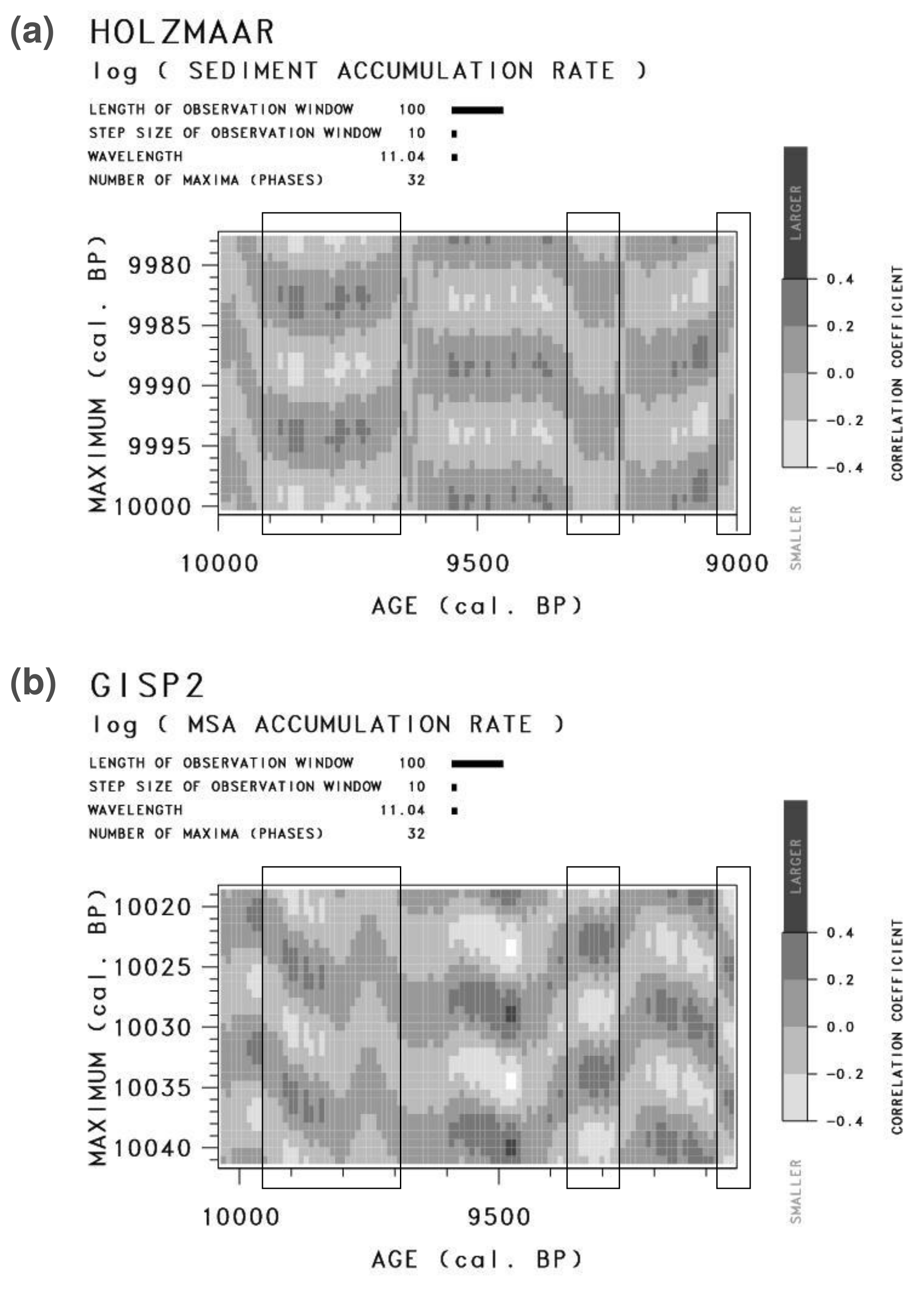}
\caption{Reproduction of Figure~17.7 of Vos et al. (2004).
Comparative phase analysis of varve thickness in Lake Holzmaar 
and MSA (Methanesulfonate)
influx measured in the Greenland ice core of GISP2. (a) 
Phase diagram of varve thickness data of Lake Holzmaar between
10000 and 9000 cal. BP. (b) Phase diagram of MSA influx data from 
GISP2 between 10041 and 9041 cal. BP.
Reprinted by permission from: Springer,  Vos, H., Br\"uchmann, C., L\"ucke, A., 
  Negendank, J.F.W., Schleser, G.H. \& Zolitschka 2004, 
  Climate in Historical Times: Towards a Synthesis of Holocene Proxy Data 
  and Climate Models. H. Fischer, T. Kumke, G. 
  Lohmann, G. Floser, H. Miller \& H. von Storch (Eds.),
  GKSS School of Environmental Research, p. 293.
  Copyright (2004).}
\label{fig1}
\end{figure}

\begin{figure}
\includegraphics[width=\linewidth]{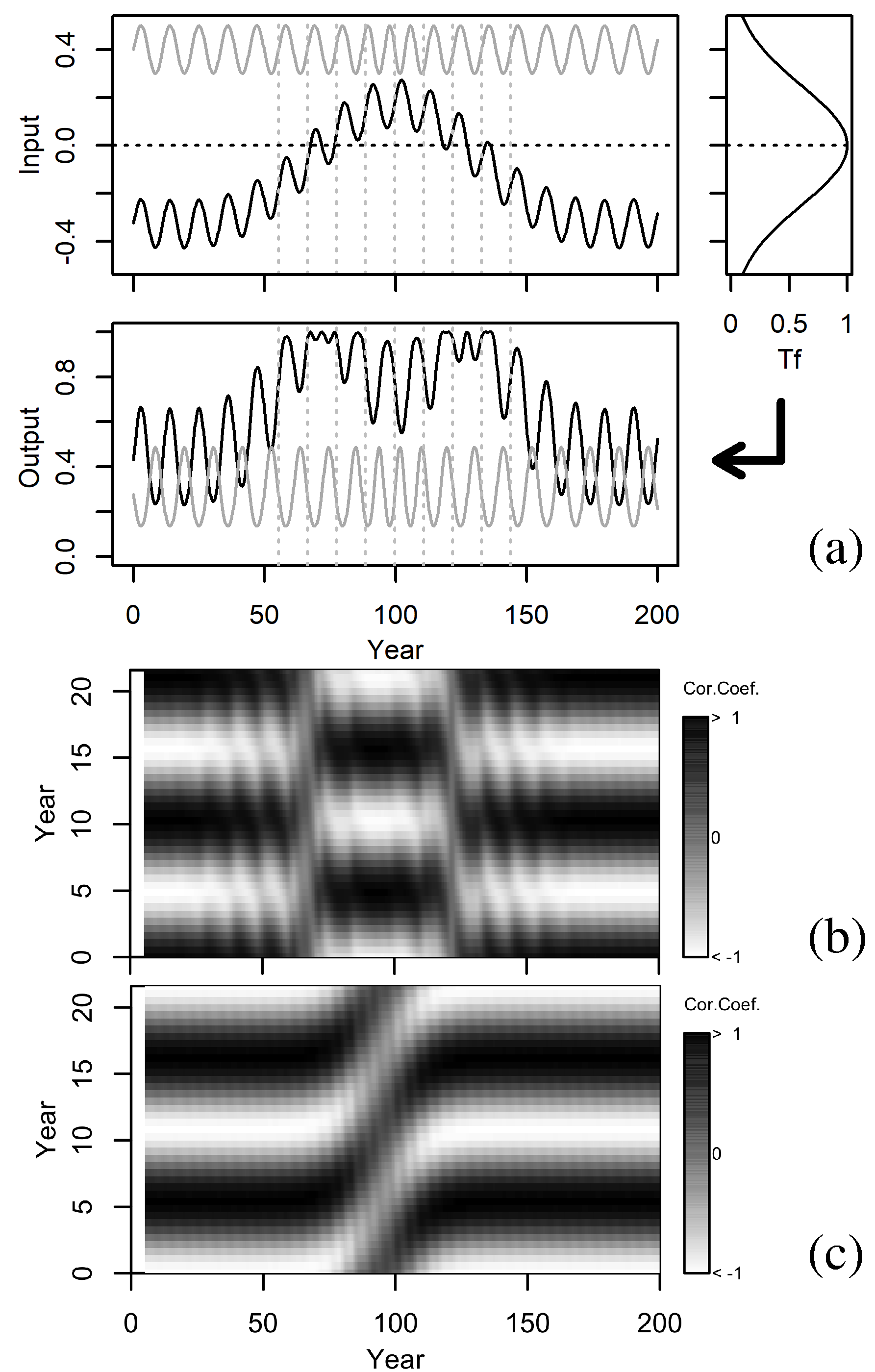}
\caption{Illustration of the effect of a transfer function on 
signals without and with real phase jumps. (a) Two input functions,
one (black) with a crossing of the maximum of the transfer function 
(right), the other one (gray) with an additional cycle around 100 
years, are transferred into corresponding output signals. 
(b) Phase diagram
of the output signal of the black curve from (a), with two
apparent phase jumps by half a cycle. (c) Phase diagram
of the output signal of the gray curve from (a), with one
real phase jump by a full cycle. 
}
\label{fig2}
\end{figure}

As carefully worked out by Vos et al. (2004),
such phase jumps by half a cycle 
should not be attributed to the underlying solar 
signal. Instead, they  appear as artifacts of applying a nonlinear 
transfer function to the input signal related to 
solar activity. This  
transfer function reflects the optimality conditions 
of the growth
of algae on abiotic input such as temperature or 
UV radiation\footnote{Fostered by the seminal paper of 
Charlson et al. (1987), the possible climatic 
effect of sulfuric
cloud-condensation nuclei produced by algae in sea water 
has been discussed intensely and 
controversially (Quinn and Bates 2011).}.
Adapting Figure~17.3 of
Vos et al. (2004), the specific convolution effect of applying 
a transfer function to a signal comprising a high frequency 
and a secular trend 
is visualized here in Figure~2. 
We add to this, however, an 
illustration of the effect of a hypothetical 
{\it real} phase jump 
in the input data (as will be discussed 
further below) on the phase diagram.
While such real phase jumps by 11 years 
in the input data would produce equivalent
11 years jumps in the phase diagram, this is not 
what was observed in Figure~1 (see, however, our
discussion in Appendix A
on the possibility that such fast phase jumps as 
in Figure~2c might have been missed
in Figure~1 due to the long comparison 
window used 
by Vos et al. (2004)). At any rate, it seems 
quite plausible that the solar cycle was  
phase coherent over the central segment 
9650-9350 cal. BP (if not the 
entire period 10000-9000 cal. BP)
with a period of 11.04 years as estimated by 
Vos et al. (2004)). Given the limited accuracy and
duration of the underlying datasets, we consider these
11.04 years as hardly distinguishable from the 
11.07-year period to be discussed in the following
(for the 90 cycles shown in Figure~1, this difference 
would result in a shift along the ordinate axis by 2.7 
years).

\section{Schove's data and  two ``lost cycles''}

We turn now to the Schwabe cycle during the last six centuries.
Anticipating the limited availability of {\it annual}
$^{14}$C and $^{10}$Be data (starting in the 16th and 15th 
century, respectively)
we restrict the considered solar cycle maxima to those 
from A.D.~1404 onward (see Table~1).
For the interval 1404 - 1755, we 
use the values from Table~2 and Appendix~B of
Schove (1983), whilst 
all later maxima (starting from 1761.5) are taken 
from the more contemporary 
Table~1 of  Hathaway (2015) (they mainly coincide with 
Schove's data, so that we indicate this series by the 
notion ``Schove''). 
The maximum  for the last cycle SC 24 has 
been roughly estimated to be 2013.0,  
which lies between the two activity peaks at the end 
of 2011 and the beginning of 2014. Basically, we utilize 
Wolf's numbering scheme for the solar cycles, 
while possible candidates of ``lost cycles'' are indicated 
by half-integers, e.g. ``-16.5'' and ``4.5'', the latter 
one  corresponding to Usoskin's notion ``SC 4' ''
(Usoskin 2002). 

\begin{table}
% \centering%%%
\caption{Maxima of solar cycles according to different sources. 
The two ``lost cycles'', as discussed by Link (1978) and  
Usoskin et al. (2002), 
are included in the columns ``Li/Us''.}
\label{table1}
\begin{tabular}{lllllll}\hline
SC & Schove & Li/Us & $^{14}$C &Li/Us&  $^{10}$Be & Li/Us\\
\hline
-31& 1404 && & &&\\
-30& 1416 && &  &&\\
-29&1428 && & &&\\
-28&1439 && & &1444&\\
-27&1450 && & &1454&\\
-26&1460 && & &1464&\\
-25&1474 && & &1471&\\
-24&1485 && & & 1481&\\
-23&1493 &&  &&1494&\\
-22&1506.5 &&  &&1503&\\
-21&1517.9 && 1524 &&1516&\\
-20&1528.2 && 1532 &&1524&\\
-19&1537.6 && 1541 &&1538&\\
-18&1547.4 && 1551 &&1552&\\
-17&1558.3 &1555& 1558 &&1558&\\   
-16.5&&1563&& 1566&&1567\\
-16&1571.3 &1570& 1576 &&1567&1577\\ 
-15&1581.5 &1581& 1584 &&1582&\\
-14&1593.8 && 1593 &&1592&\\
-13&1604.4 && 1603 &&1603&\\
-12&1614.3 && 1613 &&1614&\\
-11&1625.8 && 1626 &&1629&\\
-10&1639.3 && 1638 &&1644&\\
-9&1650.8 && 1646 &&1652&\\
-8&1661.0 && 1655 &&1660&\\
-7&1673.5 && 1664 &&1668&\\
-6&1685.0 && 1675 &&1678&\\
-5&1694.5 && 1690 &&1689&\\
-4&1705.5 && 1704 &&1705&\\
-3&1718.2 && 1719 &&1719&\\
-2&1727.5 && 1730 &&1731&\\
-1&1738.7 && 1740 &&1741&\\
0&1750.3 && 1749 &&1751&\\
1&1761.5 && 1762 &&1758&\\
2&1769.75 && 1769 &&1765&\\
3&1778.42 && 1781 &&1778&\\
4&1788.17 &1788.4& 1791 &&1789&\\
4.5&& 1795 &  &1803&&1801\\
5&1805.17 &1802.5& 1803 &1812&1801&1812\\
6&1816.42 &1817.1&  1821 &&1820&\\
7&1829.92 && 1830 &&1827&\\
8&1837.25 && 1837 &&1837&\\
9&1848.17 && 1852 &&1850&\\
10&1860.17 && 1860 &&1861&\\
11&1870.67 && 1870 &&1872&\\
12&1884 && 1886 &&1886&\\
13&1894.08 &&1895 &&1897&\\
14&1906.17 && 1906 &&1907&\\
15&1917.67 && 1918 &&1918&\\
16&1928.33 && 1927 &&1927&\\
17&1937.33 && 1938 &&1938&\\
18&1947.42 && 1948 &&1948&\\
19&1958.25 && 1959 &&1959&\\
20&1968.92 && 1970 &&&\\
21&1980 && 1982 &&&\\
22&1989.58 &&  &&&\\
23&2000.33 &&  &&&\\
24&2013 &&  &&&\\
\hline
\end{tabular}
\end{table}

The open squares in Figure~3 show the residuals 
of the times of the cycle maxima from a 
linear function with a hypothetical 11.07-year 
trend (this representation corresponds to the so-called 
``O$-$C'', i.e. ``Observed minus Calculated'' method, 
see Richards et al. 2009). Obviously, 
those residuals
form a rather horizontal band, with the 
well-known tendency for shorter (and stronger) 
cycles in the 20th century. Particularly remarkable
is also the accumulation of shorter cycles before the Dalton 
minimum, in combination with the following extremely long 
cycle SC 4, which brought the sequence of cycles back to 
the basic 11.07 periodicity. While this jumpy 
behaviour has been 
qualified as ``great solar anomaly'' (Sonett 1983), 
``irregular phase evolution'' or ``phase catastrophe'' 
(Usoskin et al. 2002),
we support here the quite contrary interpretation
of Dicke (1978) that 
the abnormally long SC 4 simply re-establishes
the phase coherence, after the few short cycles before 
were in danger of loosing synchronization.
We note in passing that the rather weak  
SC 24 could play a similar role in re-synchronizing 
the series of cycles. 

\begin{figure}
\includegraphics[width=\linewidth]{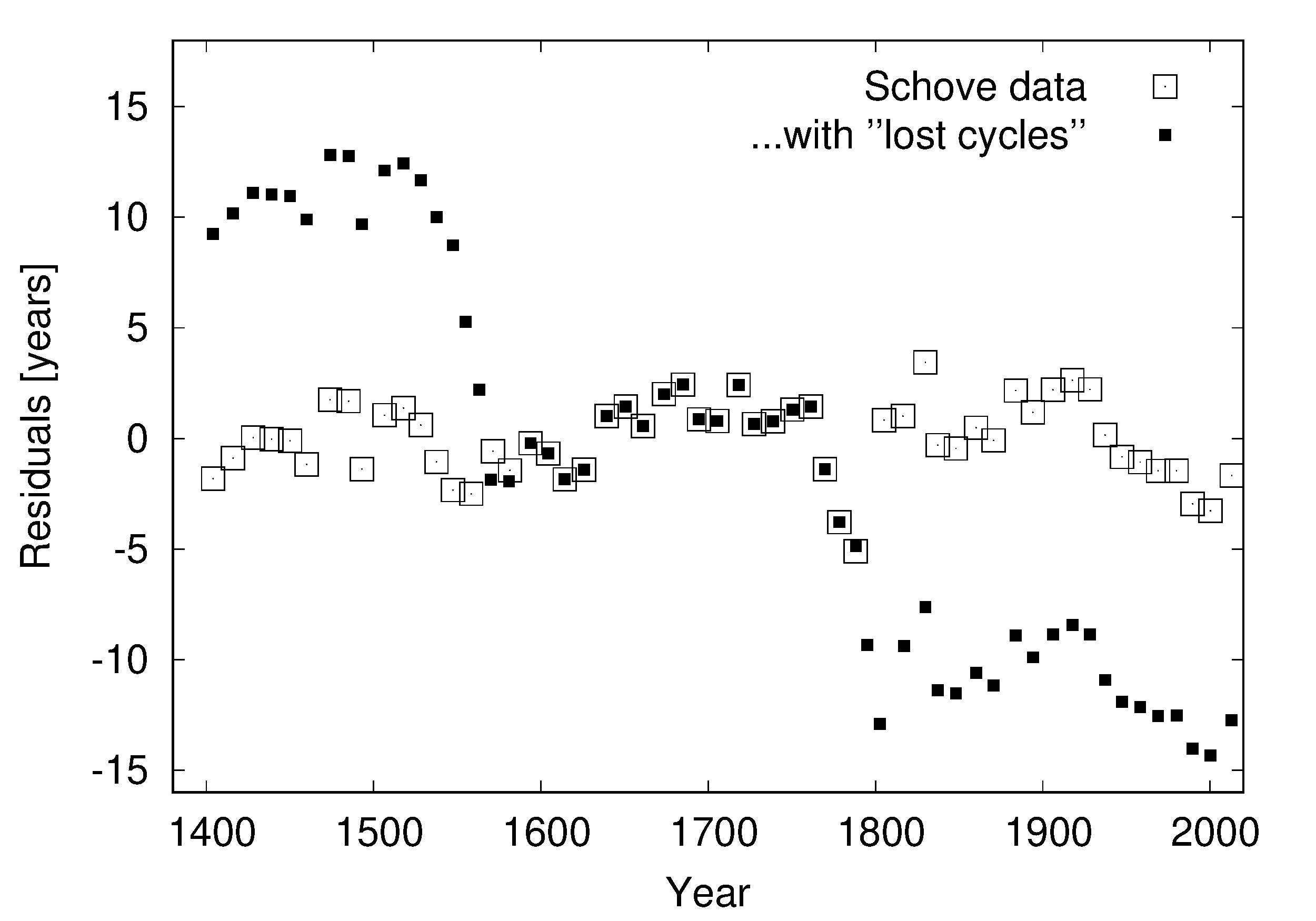}
\caption{Residuals of Schove's cycle maxima from a
linear trend with 11.07-years period, 
without (open squares) and with inclusion (full squares) 
of the ``lost cycles'' of Link (1978) and Usoskin et al. (2002). }
\label{fig3}
\end{figure}

This being said, the quest for a ``lost cycle'' within 
SC 4 is perfectly legitimate, and 
Usoskin's arguments for 
its existence were 
convincing, indeed 
(Usoskin 2002, Usoskin et al. 2009). If we add his 
additional maximum
at 1795.0, and change also the instants 
of the neighboring maxima to Usoskin's modified values, 
we obtain the full squares in Figure~3, 
showing now a sharp phase jump 
by 11.07 years.

The full squares in Figure~3 indicate also a second 
possible phase jump around 1563.
The corresponding additional maximum had been 
derived by Link (1978) from 
auroral observations, but was 
deliberately omitted by Schove with the following
justification: 
``An extra auroral cycle with a minimum in 1559 is 
due to the over-assiduous search for prodigies by 
Conrad Lycosthenes 
whose catalogue of prodigies ended at that time: in 
the Far East there is no corresponding minimum." 
(Schove 1979). As we will see below, when analyzing 
cosmogenic isotopes, the occurrence of a (weak) maximum
during this time is at least conceivable.

\begin{figure}
\includegraphics[width=\linewidth]{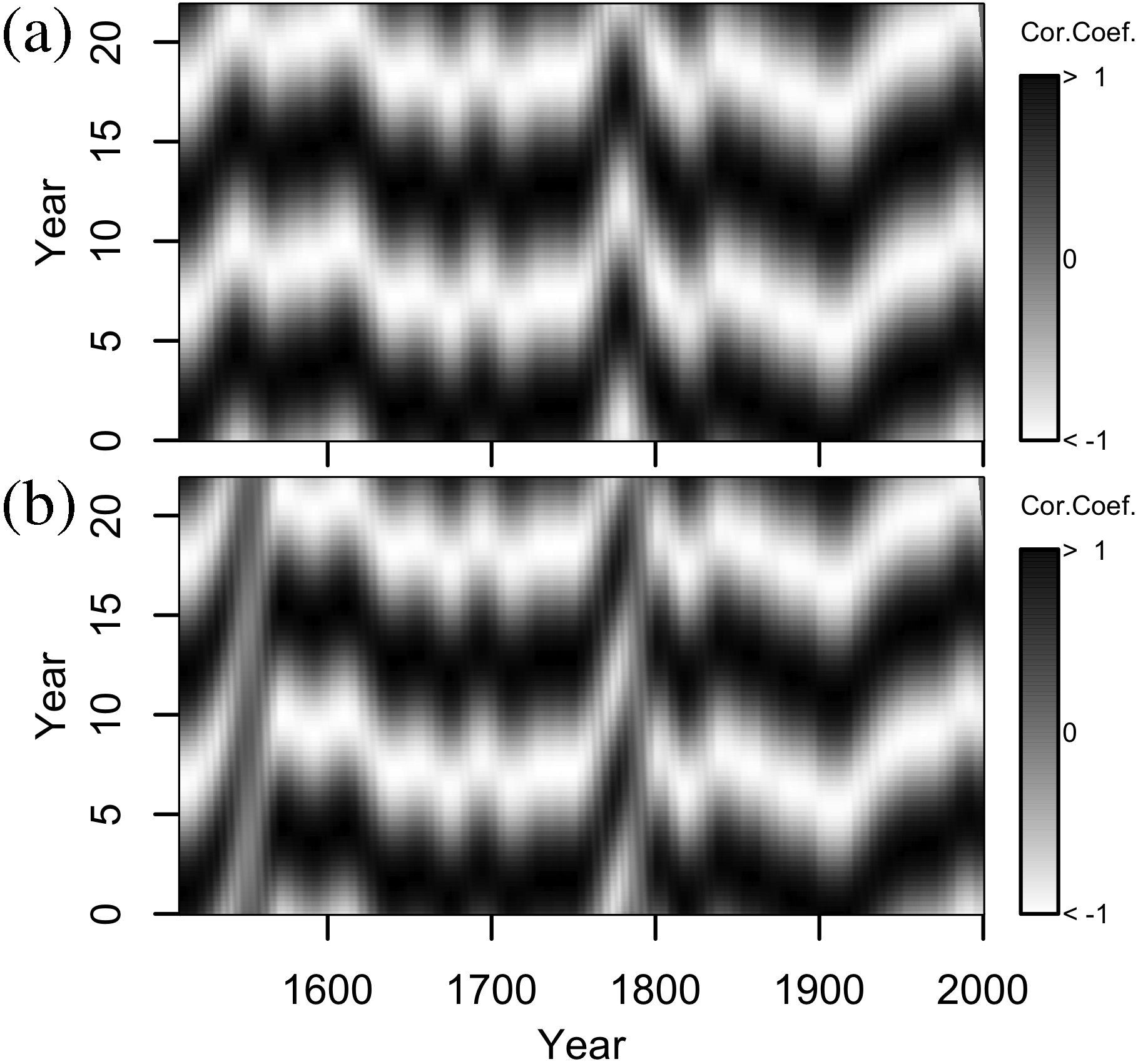}
\caption{Phase diagrams of the maxima from Table~1, 
and the corresponding minima from Table~A1, 
both for Schove's original data (a) and for the
inclusion of the ``lost cycles'' (b).
}
\label{fig4}
\end{figure}

In Figure~4 we show the phase diagrams (generated
in a similar way as Figure~17.7 of Vos et al. (2004), see Appendix A)
of these two different time series. For that purpose we enhance the 
series of maxima according to Table~1 by the corresponding 
series of minima (see Table~B1 in the appendix). 
However, since the reliability of Schove's minima 
data is compromised before 1500 we restrict Figure~4 to the 
time period 1500-2000.
Figure~4a shows the phase diagram for the pure Schove data,
in Figure~4b we have included also the two ``lost cycles''. 
The main difference between them is particularly 
visible shortly before 1800,
where Schove's data (a) exhibit a triangular shape, while the
inclusion of the ``lost cycle'' makes this region appear more like the
typical phase jump illustrated  in Figure~2c. 
The triangular shape in Figure~4a is quite similar to the one
seen in the 9800-9700 cal. BP segment 
(Figure~1b) of the GRISP2 date, 
while the corresponding segment of the Lake Holzmaar data 
(Figure~1a)
did not show such a behaviour (but see, again, our critical
discussion in Appendix A).

With  Figure~5 we would like to point out an interesting effect - or better: 
a non-effect - of phase jumps. This figure shows the most relevant segment 
(more details can be found in Stefani et al. 2020) of the
Lomb-Scargle diagram of the combined maxima and minima data, again 
for the data without and with  the ``lost cycles''.
Somewhat contrary to a naive expectation (based on a simple 
counting of maxima in a certain time interval), the dominant peak 
at 11.07 years remains nearly unaffected by the inclusion of the 
``lost cycles''. The reason for that is the {\it maintenance of 
phase coherence} over the additional cycle, so that the projection of 
the data onto a test harmonic function does barely change.

\begin{figure}
\includegraphics[width=\linewidth]{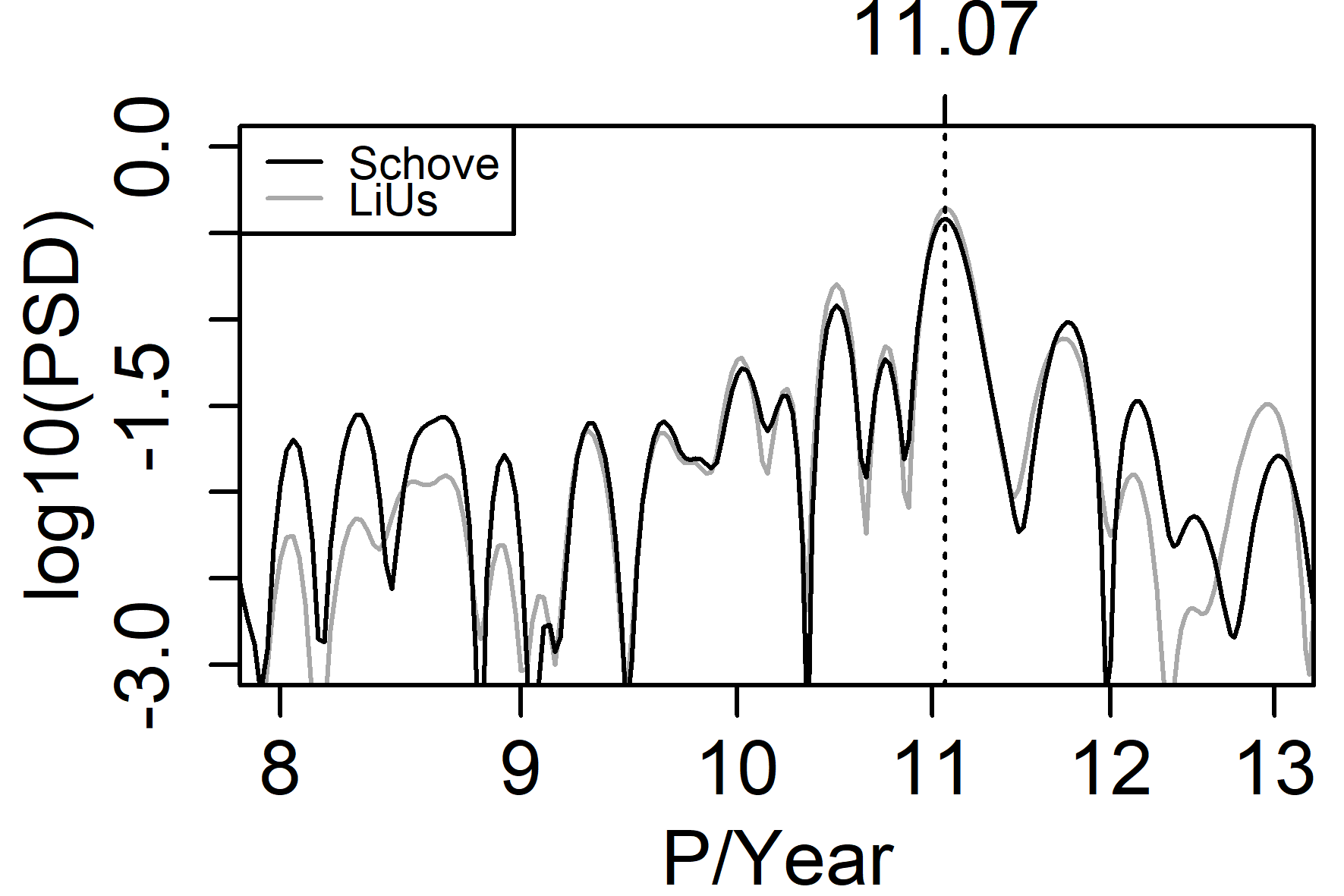}
\caption{Zoom of the Lomb-Scargle diagram for the cycle maxima and minima
corresponding to Schove's data without (black) and with (gray) 
inclusion of 
the ``lost cycles''.
For the generation of the diagram, consecutive minima 
and maxima were smoothly connected by 
segments of sine-functions.
}
\label{fig5}
\end{figure}

Note that those candidate phase jumps  
around 1565 and 1795 
do in no way contradict the concept of synchronization;
similar phase jumps are actually well known from various 
frequency-locked  
systems (Pikovski et al. 2003). 
However, the presently available data do not allow 
a conclusive decision about 
their existence. In both cases the 
corresponding data (sunspot numbers and/or radioisotopes)
exhibit a rather shallow maximum 
within a comparably long cycle, and there is a certain degree of 
ambiguity on whether to count this behaviour as an additional 
cycle or not. Note that secondary peaks within one cycle 
are not untypical for the solar dynamo, and they were 
regularly produced in our tidal synchronization model
(see Figure~7 in Stefani et al. 2016).
Unfortunately, for those historical maxima around 1565 and 1795
we do not have the sign information for the solar magnetic field,
which would be required for a (final)  
distinction between just a secondary peak or a full-blown
additional cycle. Maybe some future cycles will give us the 
opportunity to clearly identify phase jumps.

At any rate, as we have seen in the PSD's of Figure~5, the 
consequences of (potential) phase jumps
would be quite limited since the
spectrum at the dominant frequency is barely changed. 
The most important point is
the re-synchronization after those ``special events''.

\section{Cycle maxima from radionuclide data}

In this section, we attempt an independent validation of
Schove's maxima data (whose early 
parts were mainly derived from aurora borealis data) 
by annually resolved data of 
cosmogenic isotopes.
Past solar activity has often been reconstructed 
using abundances of $^{14}$C or 
$^{10}$Be measured
in stratified ice cores or (ancient) tree rings. These isotopes are
constantly being produced in the Earth's atmosphere due to
interactions of oxygen and/or nitrogen with neutrons that in turn are
generated from cosmic rays penetrating the upper layers of the
atmosphere. The corresponding production rates of the isotopes depend
on the flux of the cosmic rays, which is modulated by the periodically
changing solar magnetic field.
A comprehensive modeling of the involved processes includes the impact
of secular long term variations due to changes in the Earth's magnetic
field, the typical residence time of isotopes in different layers of
the atmosphere, as well as geophysical and climatic mechanisms. The
resulting production rates proved to correlate well with the observed
sunspot numbers (Beer et al. 2012, 2018), which in turn permits 
the reconstruction of solar activity
cycle in time periods when no sunspot observations are available.

Specifically, we will use the solar modulation function $\Phi$ 
as derived by Muscheler et al. (2007), which mainly
relies on $^{14}$C data from tree-rings, and 
two $^{10}$Be data sets of ice cores from
the Dye-3 site (Beer et al. 1990) and 
the NGRIP site (Berggren et al. 2009) 
in Greenland. For both radionuclides we will outline our
data processing scheme, including the splitting into moving 
averages and remainders. Later, when comparing the maxima resulting 
from those remainders with Schove's data, we will 
discuss again the emergence of ``lost cycles'' 
as described in the previous section.

\subsection{$^{14}$C data}

We utilize the data of Muscheler et al. (2007,2008),
and concentrate on their segment 1511-1954, where the data 
were basically
inferred from the annual tree-ring $^{14}$C data
of Stuiver et al. (1998).
These were then transformed, taking into account a geomagnetic
field reconstruction and a carbon cycle model, into a solar modulation 
potential with the unit MeV.

\begin{figure}
\includegraphics[width=\linewidth]{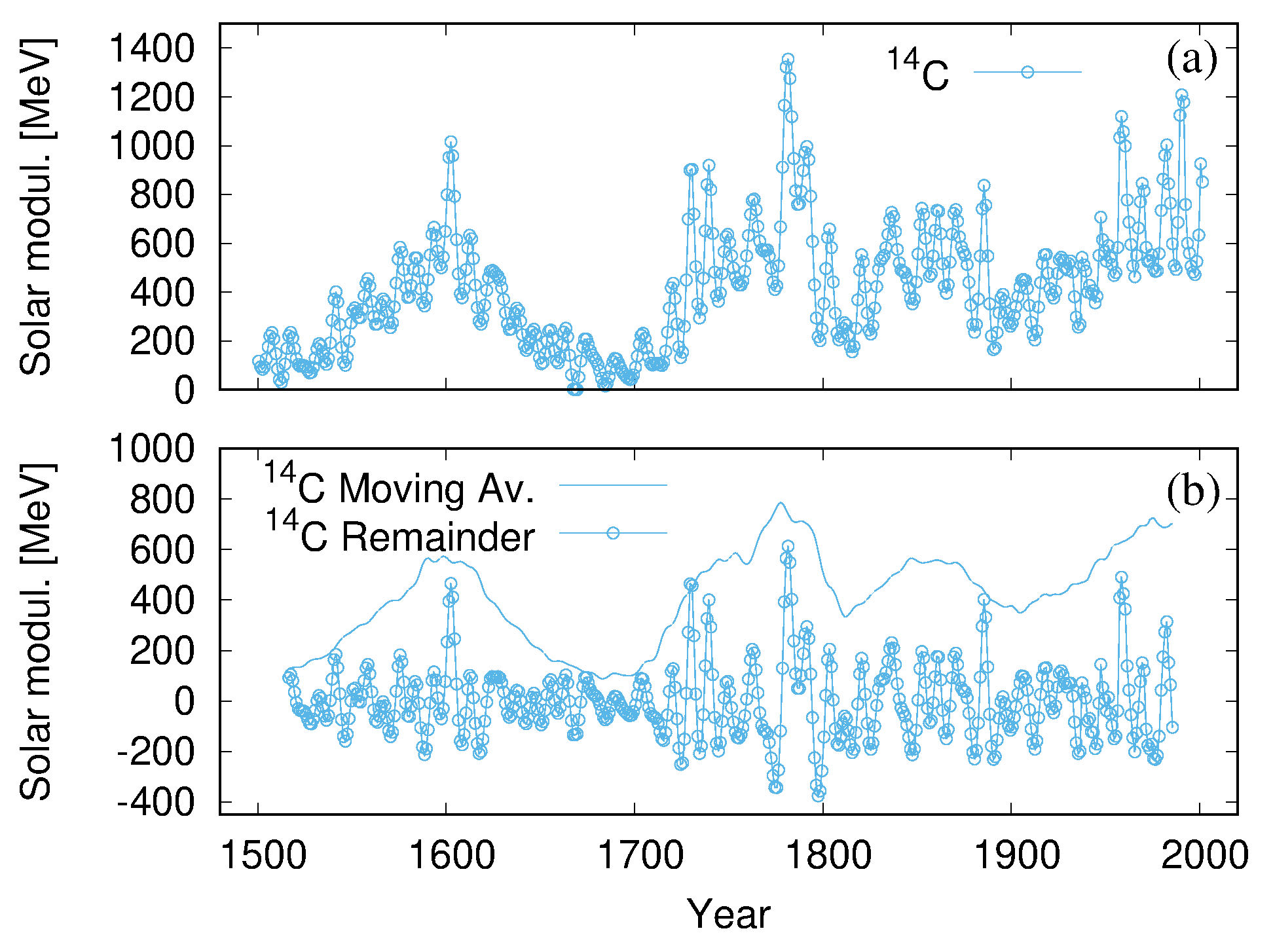}
\caption{Solar modulation potential as derived from 
$^{14}$C data by Muscheler et al. (2008) (a), and the corresponding 
moving averages and remainders (b).}
\label{fig6}
\end{figure}

Figure~6 shows the original data of Muscheler et al. (2008), together with 
their splitting into a centered moving average (with
an averaging period of 33 years) and the corresponding 
remainder. Only this remainder will be used below for the 
identification of the cycle
maxima.

\subsection{$^{10}$Be data}

The first one of the two $^{10}$Be data sets 
stems from the site Dye 3 in South Greenland,
originating from a 300-m ice core which represents 
approximately 600 years of ice accumulation. 
These data are given in a quasi-annual manner
for the time period 1424 -1986 (Beer et al. 1990). 
Specifically, in Figure~7 we show the
derived $^{10}$Be fluxes in units of atoms cm$^{-2}$ s$^{-1}$. 
Note that a specific segment between 1780-1886 had been used 
previously by Beer et al. (1990)
to evidence a good anti-correlation with the 
sunspot numbers and the geomagnetic aa index,
all dominated by a clear 11 years cycle of solar activity. 
Furthermore, the
data had also been used to prove the uninterrupted 
presence of the solar cycle
throughout the Maunder minimum (Beer et al. 1998).

\begin{figure}
\includegraphics[width=\linewidth]{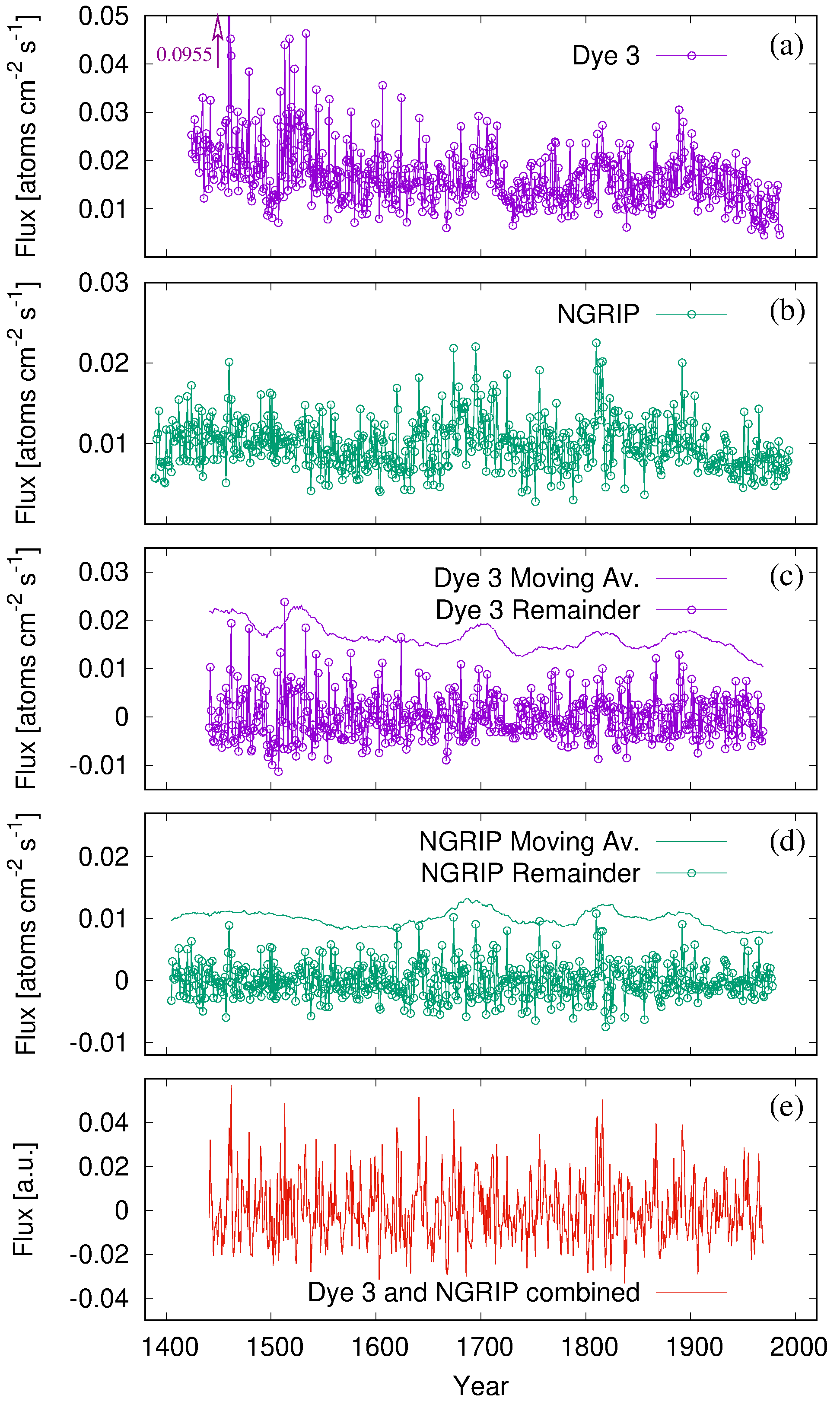}
\caption{$^{10}$Be fluxes from Dye 3 (a), NGRIP (b), their individual 
moving averages and remainders (c,d), and their combined remainders (e).}
\label{fig7}
\end{figure}

The second data set comes from the NGRIP site 
(Berggren et al. 2009), originating from an ice core of 138 m 
depth, representing the time period 
A.D.~1389-1996.

Figure~7 reveals some systematic differences 
in the fluxes at both sites, which can be attributed 
to dilution effects due to different snow 
accumulation rates and/or 
other impacts on the ice core such as remelting, shear, 
drift of ice layers, etc..

For both data, Figures~7c and 7d show the split into a 
centered moving average (with 33 years), and the 
corresponding remainders, where the exceptionally strong 
peak for Dye 3 at A.D.~1460 has already been canceled and
replaced by the average of the two neighboring data.

In order to average  (as far as this makes any sense 
with only two samples) over noise and 
geographic peculiarities of the
two sites,
in Figure~7e we combine the remainders of Dye 3 and NGRIP,
whereby the two individual contributions are weighted with 
the inverse of their standard deviation, in order to give 
them approximately equal weights. 

\subsection{Comparison of Schove's data with radioisotope data}

\begin{figure*}
\includegraphics[width=0.8\linewidth]{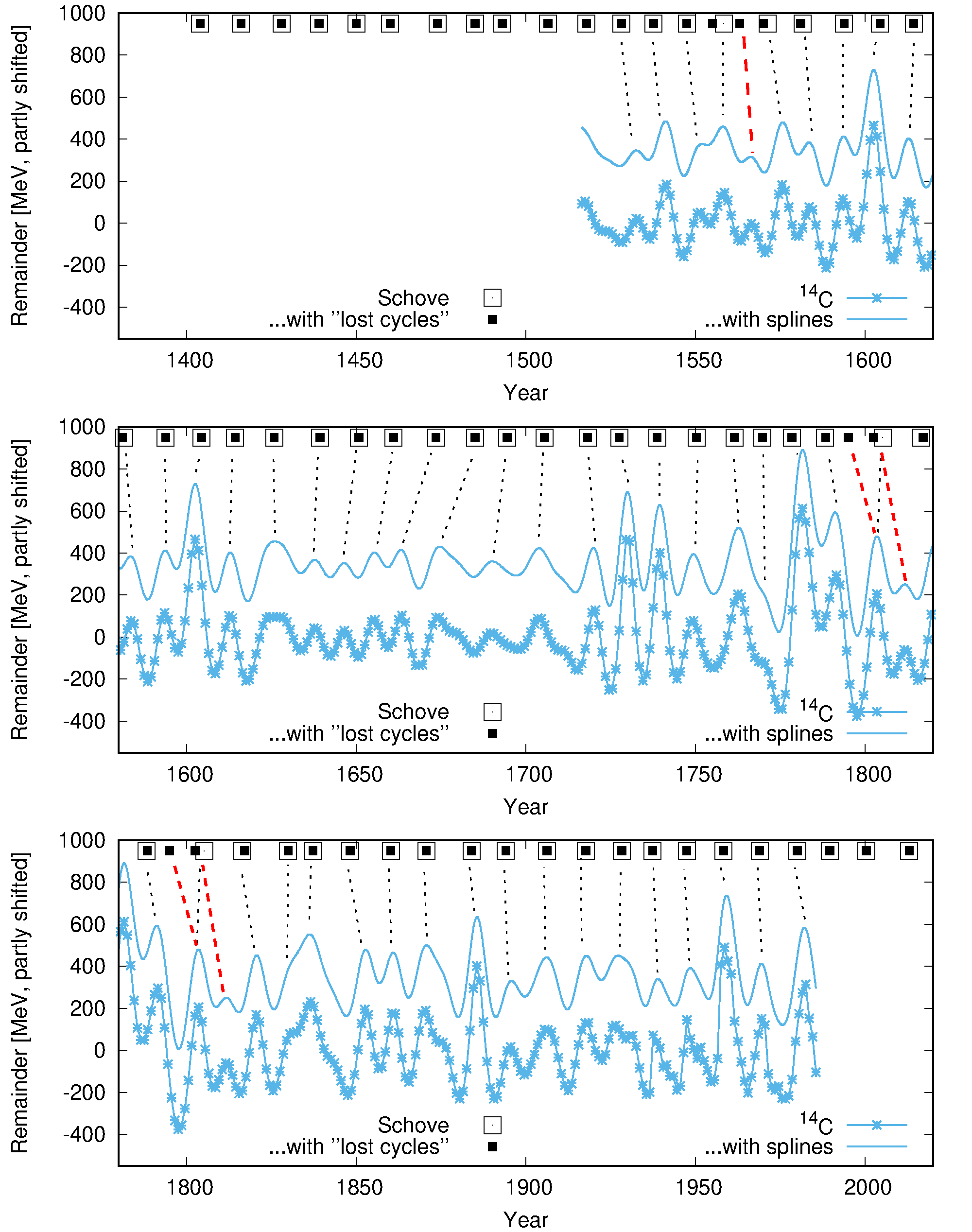}
\caption{Schove's maxima data, without (open squares) and 
with inclusion (full squares) of the ``lost cycles'', 
the remainders of $^{14}$C, and a spline interpolation thereof
(produced by gnuplot's acsplines with a weight of 
0.4).
The black dotted lines connect the $^{14}$C maxima with the maxima 
of the (conventional) Schove data. The red dashed lines denote the 
complementary relationships for the ``lost cycles''.}
\label{fig8}
\end{figure*}

In Figures~8 and 9 we present  Schove's original maxima (open squares)   
and the modified series including  ``lost cycles'' at 1563 
(full squares),
together
with the remainders of the $^{14}$C and $^{10}$Be data, respectively,
in three panels each covering an interval 
of 240 years (with some overlap). The $^{10}$Be data
were also inverted in order to make their {\it minima} 
better identifiable with Schove's {\it maxima} (the production rate 
of $^{10}$Be due to interaction of cosmic rays
in the atmosphere is reciprocal to solar activity).
The new maxima as derived from the radioisotopes 
by visual inspection are collected into the
columns 4-7 of Table~1.

Despite some differences in detail, the maxima of the 
$^{14}$C series in Figure~8 seem to be well relatable
to the corresponding maxima of Schove.
A certain exception is the time of the Maunder minimum,
in particular between 1650 and 1700,
were the $^{14}$C maxima are systematically 
shifted to earlier times when compared with Schove's maxima.
Actually, this is in very good agreement 
with the open solar flux model of Owens et al. (2012) 
which shows that for weak solar cycles the relation between the 
sunspot number $R$ and the solar modulation potential $\Phi$ 
changes
from in-phase to anti-phase.
It remains to be seen whether the pronounced hemispherical 
character of the 
solar dynamo during the Maunder 
minimum (see Figure~6 in Ribes and Nesme-Ribes 1993)
might play a complementary role in this respect.

\begin{figure*}
\includegraphics[width=0.8\linewidth]{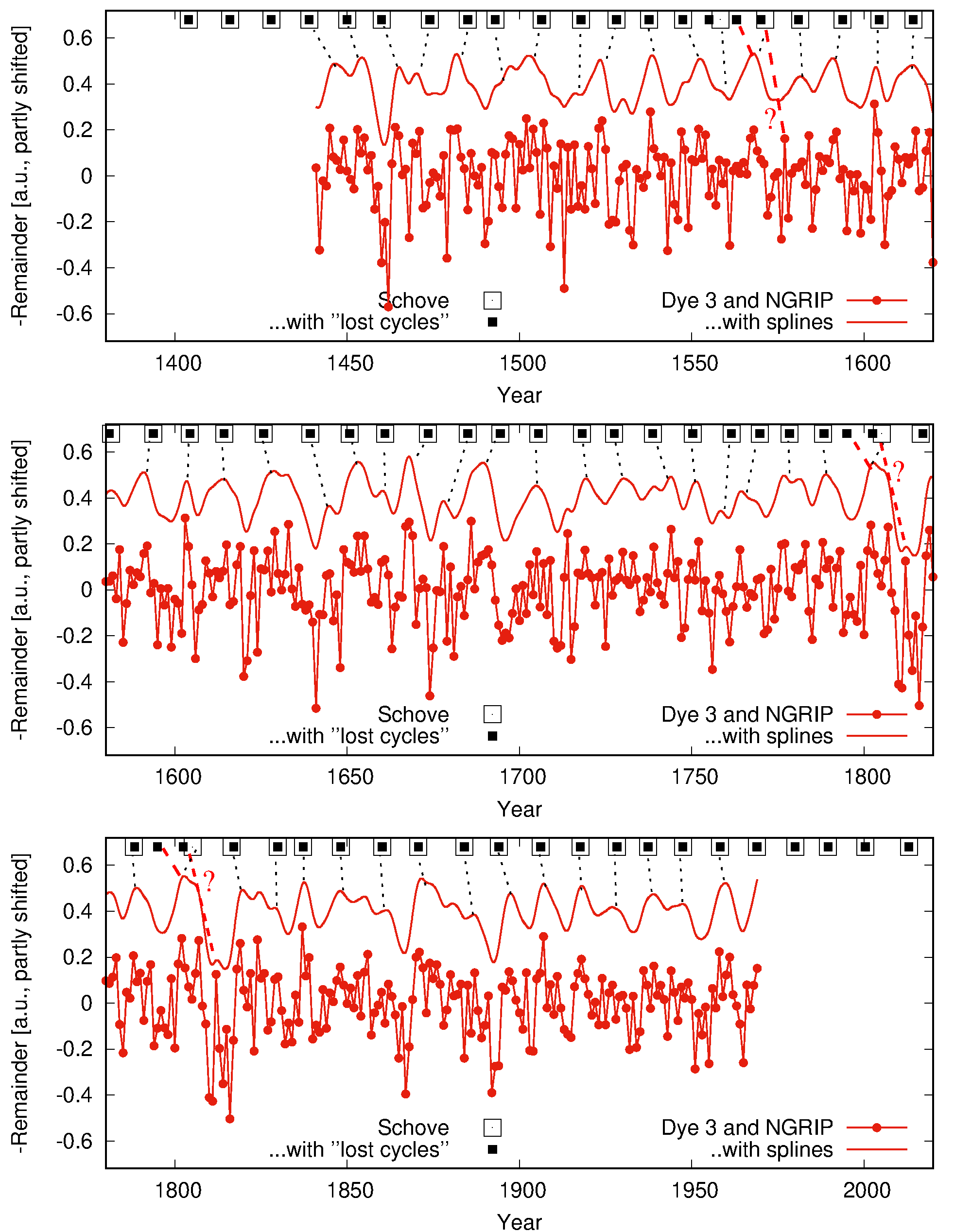}
\caption{Schove's maxima data, without (open squares) and 
with inclusion (full squares) of the ``lost cycles'', 
inverted remainders of $^{10}$Be data, and a spline interpolation 
thereof (produced by gnuplot's acsplines with a weight of 
0.4).
The black dotted lines connect the $^{10}$Be maxima with the maxima 
of the (conventional) Schove data. The red dashed lines denote the 
complementary relationships for the ``lost cycles''.}
\label{fig9}
\end{figure*}

What about the ``lost cycles''? They are indicated by the 
dashed red lines in Figure~8, and given in the fifth column
of Table~1. 
First, we can clearly 
identify an additional (weak) maximum at 1566, which 
seems relatable to Link's  additional maximum at
1563. In this sense, the $^{14}$C data would 
speak against Schove's deliberate omittance 
of this cycle. As for Usoskin's ``lost cycle'', things are 
more complicated. 
The number of cycles indicated by the $^{14}$C data
in the interval 1785-1820 signifies 
the emergence of one more maximum compared with 
Schove's data.
However, this additional maximum does not lay
at 1795, as proposed by Usoskin (2002), but 
significantly later (at 1803). Similar time shifts apply also 
to the next two maxima (1812 vs 1805 and 1821 vs 1816).
Things are even worse, since the weakest maximum, which we would 
expect to be close to Usoskin's weak maximum at 1795, 
is actually found at 1812.

As a matter of fact, we find clear deviations between 
radionuclide data and Schove's data in the interval  
1785-1820, but without obvious systematics, so that 
for the moment this puzzle cannot be solved. 
We should keep in mind that the solar modulation 
potential in itself is a complicated 
functional of the solar magnetic field, whose
topology might have changed during the ``phase catastrophe''
at the beginning of the Dalton minimum.
In view of this uncertainty, we 
will continue with the consideration of both 
possible 
series of $^{14}$C maxima, without and with 
including both ``lost cycles'', as 
summarized in the fourth and fifth column of Table~1.

Figure~9 shows the corresponding relationships for the $^{10}$Be data.
Quite generally, these data are much more noisy than the 
$^{14}$C data, so that
there is more ambiguity, first, in choosing a sensible 
smoothing parameter of the
spline interpolation and, second, in relating the resulting maxima to 
Schove's maxima. Hence, the choices made in Figure~9, and the resulting  
columns 6 and 7 in Table~1, should definitely be taken 
with a grain of salt.
As for the Maunder minimum, we see 
that the  shift to earlier
times, as discussed for the $^{14}$C data, is also
observable in the $^{10}$Be data, which gives further confidence
in the corresponding model of Owens et al. (2012). 
The situation around 1565 (indicated by a question mark) 
with the first candidate of a ``lost cycle'' is quite peculiar. 
If we relate (one black dotted line) the  clear $^{10}$Be
maximum at 1567 with Schove's maximum at 1571.3, we end up
without any ``lost cycle'' at this point. Alternatively,
we could relate the 1567 maximum with Link's maximum at 1563,
and the next (rather weak) $^{10}$Be maximum at 1577 
(which is only visible 
in the remainder, but smeared out in the splines) with 
Schove's maximum at 1571.3. With this second variant (indicated by 
two red dashed lines), we would include the ``lost cycle'' around 1565. 
A similar problem arises around 1800, although the 
qualitative behaviour of the $^{10}$Be data appears quite similar as the
corresponding behaviour of $^{14}$C discussed above. 

\begin{figure}
\includegraphics[width=\linewidth]{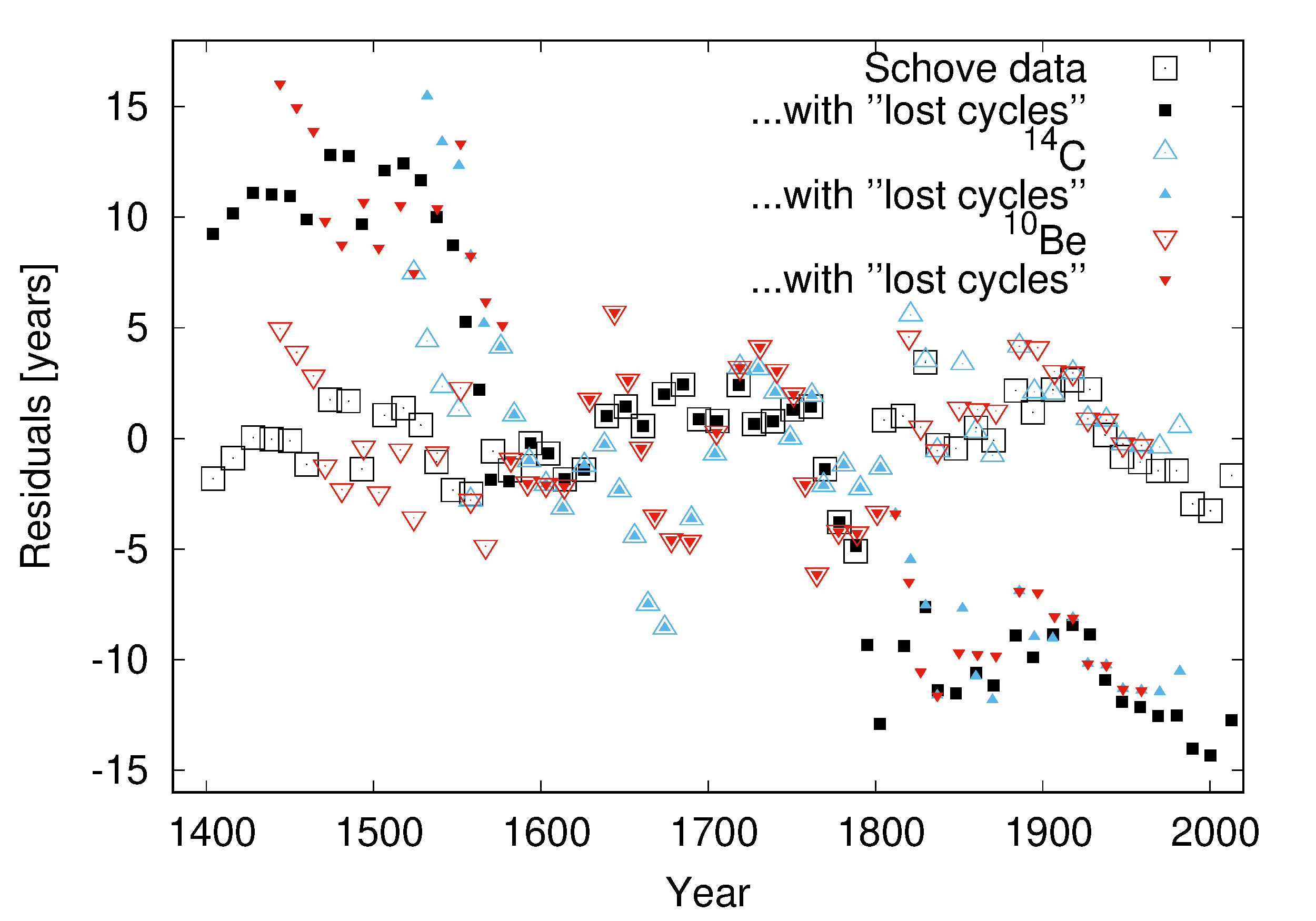}
\caption{Residuals of Schove's maxima data without and with inclusion of ``lost cycles'', 
and of the corresponding maxima of the $^{14}$C and $^{10}$Be data. }
\label{fig10}
\end{figure}

In Figure~10 we illustrate all maxima data identified so far (for details 
see Table~1). Besides a re-plot (from Figure~3) of the two variants of 
Schove's data we include now the maxima
derived from $^{14}$C and $^{10}$Be, also in the two variants
without or with the  ``lost cycles''. In general, we see 
reasonable agreement between the various data sets. The 
conspicuous ``downswing'' of both radioisotope data 
in the Maunder minimum interval 1650-1700
can consistently be explained by the
phase-shift model of Owens et al (2012).

\begin{figure}
\includegraphics[width=\linewidth]{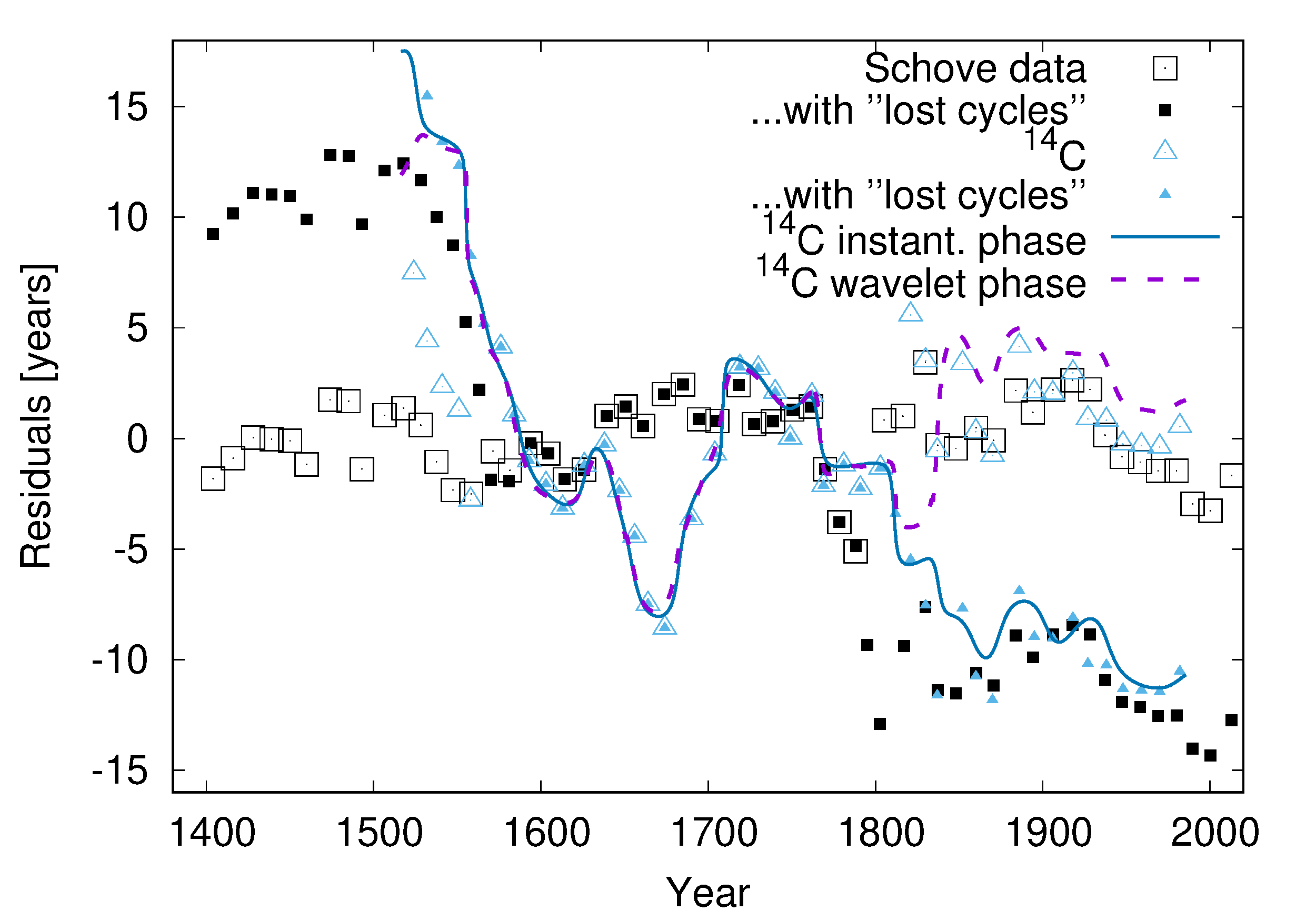}
\caption{Residuals of Schove's maxima
 without and with inclusion of ``lost cycles'',
and of the corresponding phase drifts (transformed into residuals) based on 
the instantaneous phase using a bandpass (between 
17.85-year and 8-year periods) and a wavelet analysis. 
Note the divergence of the two phase drift curves after 1800.}
\label{fig11}
\end{figure}

Figure~11 is a modified version of Figure~10, with the 
less reliable $^{10}$Be maxima data omitted, but 
a deeper analysis of the
$^{14}$C  data. The residuals of their
individual maxima are complemented 
by two continuous curves resulting from two types of 
phase drift computations.
The first one (full line) is based on determining the instantaneous phase 
from band-passed data (Seilmeyer and Ratajczak 2017). The second one (dashed line) is based on a wavelet
analyses (for details, see Appendix A). While 
both curves are nearly identical in the early segment
of the data (where they cling to the maxima including Link's 
``lost cycle''), they suddenly diverge after 1800. The curve 
based on the instantaneous phase continues according to the inclusion of 
Usoskin's ``lost cycle'', while the wavelet based curve jumps back to the
$^{14}$C  maxima data without ``lost cycle''. Actually, the final results of both 
phase drift methods depend on details of their numerical implementation
(width of the bandpass, wavelet parameter, etc.); at the 
critical points, where the amplitude of 
the 11.07-year oscillation
drops significantly, they
can either jump by one cycle or not.
Loosely speaking, these sophisticated mathematical tools
can complement, but not completely replace,  
the subjective decision (by visual inspection) 
on whether a (typically shallow) maximum  should be counted as 
an additional cycle or not.

\section{Conclusions}

This paper was concerned with two types of phase
jumps related to the solar cycle and its reflections in 
different proxy data. We started by highlighting the
extremely important, but widely overlooked, results of 
Vos et al. (2004) which suggested, both for 
varved sediments from lake 
Holzmaar and for MSA distributions in the GISP2 ice core, 
an amazingly phase-coherent 
solar signal with a period of 11.04 years
over the considered 1000 years period in the early Holocene.
Then, 
we had a closer look into the ``lost cycle'' 
at the beginning of the Dalton minimum. While we remain rather 
indifferent with respect to it being real or not, in contrast to 
Usoskin (2002) we would interpret its existence - rather than its
non-existence - as a ``phase catastrophe'', or better, a 
phase jump within the otherwise synchronized Schwabe cycle 
with its 11.07 years periodicity. The same classification would
apply to a similar ``lost cycle'' around 1565, which had been
deliberately disregarded by Schove, despite some evidence 
in auroral data as pointed out by Link (1978). 
Finally, we have 
analyzed in some detail a series of $^{14}$C related data,
and two series of $^{10}$Be data. After subtracting moving 
averages, the remainders of both data exhibit signals 
with a dominant 11  years periodicity, with the $^{14}$C data 
giving significantly clearer results. When comparing these 
data with Schove's historical data we found - besides
a reasonable agreement for most of the cycle maxima -
some evidence for both Usoskin's ``lost cycle'' around 1800 and
for Link's ``lost cycle'' around 1565.
A careful re-analysis of the MSA distribution in 
the GISP2 ice core (Appendix A)
suggests, in turn, that 
corresponding fast phase
jumps in the early Holocene period might be not so 
strictly excluded as could be reasoned from the 
original work of Vos et al. (2004), who had used a comparably
long observation window which suppresses short term events.

We would like to point out once more
that the existence of phase jumps
should in no way be considered a counter-argument against 
synchronization. To the contrary: their comparably sharp 
appearance within much longer phase-coherent intervals 
speaks even {\it in favour} of an underlying synchronization.
However, as long as we have no sign information about the
solar field during these times, there will always remain some 
ambiguity in attributing the (typically shallow) 
maximum to an additional full cycle (or not). 
Fortunately, this counting problem has only limited consequences, 
since the dynamo, having gone through this  
``nervous episode'', settles again into an 
ordered, synchronized state. 
Even the PSD of the cycle time series seems to be rather unaffected by 
those events, since the dominant cycle continues with the 
same phase relation.

Finally, we would like to discuss another interesting aspect of 
phase jumps. Regardless of their very existence (or not), 
Figure~3 exhibits a tendency for a systematic shortening of the
Schwabe cycle approximately every 200 years. This mirrors 
the well-known Suess-de Vries cycle, which we had also 
confirmed in analyzing the spectrum of the residuals 
(Stefani et al. 2020). While much effort had been spent
to explain this (and other) long-term cycle(s)
in terms of corresponding periods of planetary tidal torques 
(Abreu et al. 2012), we pursued and extended the 
somewhat different idea of Cole (1973), Solheim (2013), 
Wilson (2013) that those long-term cycles may also arise as
{\it beat periods} between significantly shorter cycles.
Specifically, we enhanced our tidally synchronized 
solar dynamo model (which yields the 11.07-years Schwabe cycle) 
by a modulation of the field storage capacity of the 
tachocline with the period 
of the orbital angular momentum of the Sun which is dominated 
by the 19.86-year Jupiter-Saturn heliocentric 
conjunction period\footnote{Frankly admitting  that  the  specific 
coupling mechanism of the 
orbital angular momentum of the Sun around the solar system barycenter 
into some dynamo relevant flow parameters remains unclear, 
a modification of the very sensitive adiabaticity in the tachocline
by some sort of spin-orbit coupling seems, at least, not 
completely unrealistic}.
This modulation of the 22.14-years Hale cycle provided
a 193-year beat period of dynamo activity, which is indeed 
close to the Suess-de Vries cycle. 
With  stronger 19.86-year forcing, we observed even
additional Gleissberg-type 
periodicities around 100 years, which were 
intimately connected with
the emergence of a Wilson gap, i.e. the prevalence of 
side bands around the 11.07-years cycle
both at too short (e.g. 10 years) and too long (12 years) 
periods. The occurrence of such side bands 
can be explained by assuming that, intermittently, 
the shorter driving from spin-orbit coupling 
takes over and enslaves the mean Schwabe cycle
to $19.86/2=9.93$ years, an effect which later must be compensated 
by one or a few longer cycles to ``keep pace'' with the basic 11.07-years 
tidal forcing cycle (which is still believed to be dominant).
Yet, if the accumulated shortening of the residuals
has become too extreme (see the 
-5 years residual around 1790 in Figure~3),
then the next dynamo cycle might alternatively 
``slip back'' by one tidal driving cycle.
This would exactly correspond to the sort of  phase jumps 
that we have been discussing in this paper.

While the long phase coherent period in the 
early Holocene together with
the detailed analysis of the Schwabe cycle during the 
last 600 years has  lend 
greater plausibility to our 
starting hypothesis of a tidally synchronized 
solar dynamo, we would like to encourage 
more investigations into the 
Schwabe cycle during other periods.
Given the amazing quality of the varve thickness and
MSA data as documented by Vos et al. (2004), 
it seems highly promising to investigate those data 
also for other periods in the Holocene.

\acknowledgements
This project has received funding
from the European Research Council (ERC) under the
European Union's Horizon 2020 research and innovation programme
(grant agreement No 787544).
It was also supported in frame of the Helmholtz - RSF
Joint Research Group ``Magnetohydrodynamic instabilities'',
contract No HRSF-0044 and RSF-18-41-06201.
We are grateful to Antonio Ferriz Mas, Peter Frick, 
Rafael Rebolo, 
G\"unther R\"udiger, Dmitry Sokoloff, Willie Soon and 
Steve Tobias for fruitful discussions on various aspects of 
the solar dynamo. We would like to thank Heinz Vos and 
Bernd Zolitschka 
for comments on an early draft of this paper, and for
providing us with Figure~1.

% Example of using BiBTeX (plus natbib):
% For details see \cite{1999MNRAS.309..731B},
% \cite{1893PASP....5..204C},
% \cite{2008IAUS..252...75L}. It has been demonstrated that this
% is important \citep{2012AN....333..663S}.

% Use this code if you wish to generate your bibliography with BibTeX;
% please replace first the string "an-demo" below with the name(s) of
% the BibTeX data base(s) you want to use.
% The resulting bibliography-output (the contents of the .bbl file)
% must be pasted into this file before submission.
% 
%\bibliographystyle{an}
%\bibliography{an-demo}
% 
% Replace the following example bibliography with your references
% before submission:

\appendix

\section{Computational methods: phase diagrams and wavelet analysis}
In this appendix we present the numerical methods as they were used 
for producing the phase diagrams and for  the wavelet analyses.

\begin{figure}
\includegraphics[width=\linewidth]{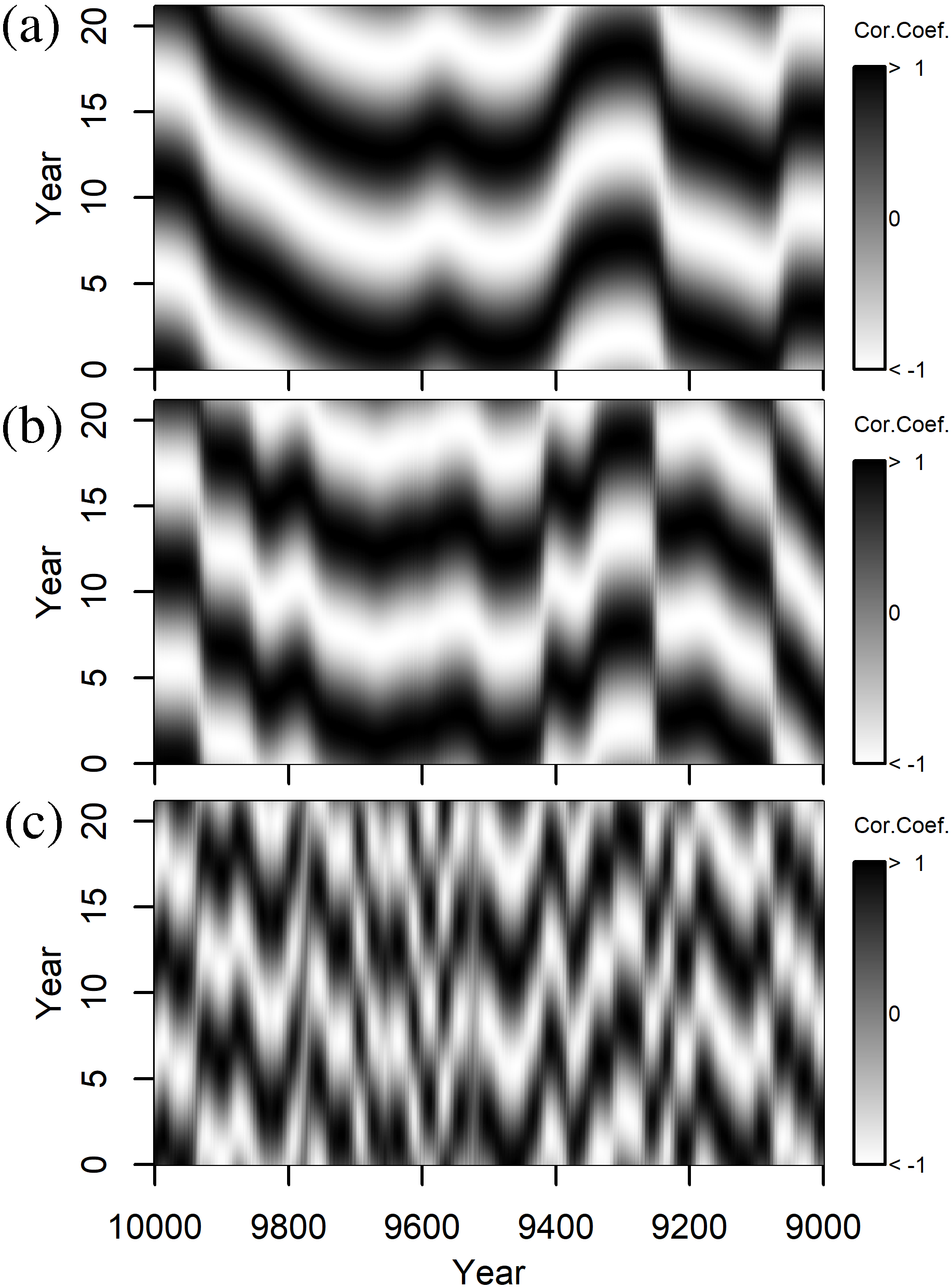}
\caption{Phase diagrams of the MSA data from GISP2, as 
generated by means of the phaselet method with bandwidths
$b=$(150~yr)$^{-1}$ (a), 
(85~yr)$^{-1}$ (b), (29~yr)$^{-1}$ (c). Not surprisingly,
(b) yields a similar result as the method of Vos et al. (2004)
with their rectangular time-domain window over 100 years, 
shown in Figure~1b. While the narrow bandwidth (a) gives a 
very smooth result, the wide bandwidth (c) suggests the occurrence of 
more transitions, which even might include real phase jumps by 11 years. 
}
\label{figaa1}
\end{figure}

\subsection{Phase diagrams}
For the generation of the phase diagrams, we assume a 
time series to be given as an equally spaced interpolation on a 
regular grid. To these data we apply a 
band-pass filter, with a central frequency $f_{0}=$(11.07 yr)$^{-1}$, 
that consists of a weight function
\begin{eqnarray}
w(f,f_{0},b)=\begin{cases}
1-\left|   \frac{f-f_{0}}{b}     \right|^{3}  &,\left|f-f_{0}  \right|  <  b\\
0  &,\mathrm{else}
\end{cases}
\end{eqnarray}
with cubic slope and bandwidth $b$.
Given the Fourier transform of the signal to be
\begin{eqnarray}
S(f)=\mathcal{F}(s(t))
\end{eqnarray}
the band pass filtered spectrum is defined as
\begin{eqnarray}
S_{\mathrm{SP}}(f)=w(f,f_{0},b)\cdot S(f)
\end{eqnarray}

The shape of the bandpass filter in the frequency domain 
causes a specific shape of the observation window and vice 
versa. The selected cubic form corresponds to an overdamped 
sinc-like observation window including the main peak and 
only the first side peaks. Vos et al. (2004) introduced a 
rectangular window in time domain (ideal masking) which 
reflects an ideal sinc kernel in the frequency domain leading 
to undesired side bands. The latter bring about an unnecessarily
wide observation window, which reduces the side bands 
but looses details in the resulting phase diagram.
According to signal theory, the bandwidth is limited to $b<0.5 f_{0}$ 
(violating this would lead to distortions due to aliasing).
Note that an observation window of 100 years in time domain, 
as used by Vos et al. (2004),
corresponds roughly to a bandwidth of $b \approx$(85~yr)$^{-1}$ in our case
of a cubic frequency window.

Now we set up a function for the ``phaselet''
\begin{eqnarray}
\phi(t,\varphi,f_{0})=\begin{cases}
\sin\left(2\pi f_{0}(\varphi+t)\right) & -1/f_{0}\leq\varphi<1/f_{0}\\
0 & \mathrm{else}
\end{cases}
\end{eqnarray}
where $\varphi$ has the meaning of the shift (along the ordinate axis 
in the phase diagrams) and $t$ selects the time position in the data.
The correlation is carried out in the usual way, so that the phaselet analysis
\begin{eqnarray}
\mathcal{P}(t,\varphi,f_{0})=\int_{-1/f_{0}}^{1/f_{0}}s(t)^{*}\phi(t+t',\varphi,f_{0})\thinspace\mathrm{dt'}
\end{eqnarray}
can be given in form of the common correlation integral in the limits
of $[-f_{0}^{-1},f_{0}^{-1}]$. Doing this for all $t$ and $\varphi$
results in the phase diagrams as presented in this paper.

As an example, we re-analyze with our method the MSA data
from ice core GISP2 (Saltzmann et al. 1997) 
in order to compare the results with those of Vos et al. (2004),
re-plotted here in Figure~1b. 
Figure~A1 shows the resulting phase diagrams for a variety of 
bandwidths $b=$(150 yr)$^{-1}$ (a), (100 yr)$^{-1}$ (b), (85 yr)$^{-1}$ (c), and 
(50 yr)$^{-1}$ (d). While our result for $b=85$ (yr)$^{-1}$
corresponds approximately with Figure~1b, we can either
get smoother results (for the narrow bandwidth (150 yr)$^{-1}$), 
or strongly fluctuating results (for the narrow bandwidth (50 yr)$^{-1}$).

\subsection{Wavelet analysis}

The spectral properties of a signal at a certain time can be 
conveniently revealed by wavelet analysis. 
The continuous wavelet transform of the real signal $f(t)$ is 
defined as
\begin{equation}
w_f(\tau,t)={\tau}^{-1}\int_{-\infty}^{\infty}f(t)\psi^*\left(\frac{t'-t}{\tau}\right)\mathrm{dt'},
\label{wF}
\end{equation}
where $\psi$ is the analyzing wavelet function, $\tau$ and $t$ define the scale 
(period of oscillation) and  time, respectively. 
The specific choice of $\psi$ is always a compromise to balance resolution in 
time and scale. 
The popular Morlet wavelet 
\begin{equation}
  \psi(t)=e^{-t^2} \left(e^{ 2\pi i t  }-e^{- \pi ^2 }\right)
\end{equation}
is used to satisfy the most of requirements for processing 
various records of solar activity (Frick et al. 2020).
In addition to many widely used diagnostics we utilize 
wavelets here to perform the phase analysis. The phase drift 
between two oscillations with period $\tau$ in signals  
$f_1(t)$ and $f_2(t)$ evolves as
\begin{equation}
\delta\phi(t)=\arg(w_1(\tau,t)w_2^*(\tau,t)).
\label{wphase}
\end{equation}
Figure~11 shows the residuals $-\tau \delta\phi(t)/(2 \pi)$ with 
\begin{equation}
w_1=\exp(2\pi i (t - 66.75 \cdot 11.07 + 1000)/11.07)
\end{equation}
and $w_2=w_{^{14}C}(11.07,t)$. Also 
we added  $\pm2\pi$ appropriately 
to treat a discontinuity caused by the $\arg$-operation.

\section{Schove's minima data and a waterfall diagram}
In Table~B1 we complement the maxima data from Table~1 by the 
corresponding minima data which were used 
for the generation of the Lomb-Scargle diagram in Figure~5.
The restriction to the period 1500-2010 is due to the fact that 
Schove's minima data before 1500 are not of the same reliability 
as the corresponding maxima data. The reason for that is  that 
the most recent minima data in the Appendix 
of Schove (1983) were spoiled by wrong data 
between A.D.~511 - 1493: those were erroneously 
copied from the table of the corresponding maxima.

\begin{table}
% \centering%%%
\caption{Minima and maxima  
of solar cycles according to Schove (1983) and Hathaway (2015). 
The ``lost cycles'' as discussed by Link (1978) and  
Usoskin et al. (2002)  are 
included in the columns ``Li/Us''.}
\label{table2}
\begin{tabular}{lllll}\hline
&\multicolumn{2}{c}{Min}&\multicolumn{2}{c}{Max}\\
\hline
SC & Schove & Li/Us&Schove & Li/Us \\
\hline
-22&1501.5 &&1506.5 &\\
-21&1513.7 &&1517.9 &\\
-20&1524.7 &&1528.2 &\\
-19&1534.3 &&1537.6 &\\
-18&1543.7 &&1547.4 &\\
-17&1554.5 &1551.0&1558.3 &1555\\
-16.5&&1559.0&&1563\\
-16&1567.5 &1566.0&1571.3 &1570\\ 
-15&1578.2 &1576.0&1581.5 &1581\\
-14&1587.5 &&1593.8 &\\
-13&1598.8 &&1604.4 &\\
-12&1609.2 &&1614.3 &\\
-11&1620.2 &&1625.8 &\\
-10&1633.7 &&1639.3 &\\
-9&1645.5 &&1650.8 &\\
-8&1655.9 &&1661.0 &\\
-7&1666.7 &&1673.5 &\\
-6&1679.5 &&1685.0 &\\
-5&1689.5 &&1694.5 &\\
-4&1699.0 &&1705.5 &\\
-3& 1712.5&&1718.2 &\\
-2&1723.5 &&1727.5 &\\
-1&1734.0 &&1738.7 &\\
0&1755.2 &&1750.3 &\\
1&1761.5 &&1761.5 &\\
2&1766.5 &&1769.75 &\\
3&1775.5 &&1778.42 &\\
4&1784.7 &1784.3&1788.17 &1788.4\\
4.5&& 1793.1&& 1795 \\
5&1798.3 & 1799.8&1805.17 &1802.5\\
6&1810.6 &1810.8&1816.42 &1817.1\\
7&1823.3 &&1829.92 &\\
8&1833.9 &&1837.25 &\\
9&1843.5 &&1848.17 &\\
10&1856.0 &&1860.17 &\\
11&1867.2 &&1870.67 &\\
12&1878.9 &&1884 &\\
13&1889.6 &&1894.08 &\\
14&1901.7 &&1906.17 &\\
15&1913.6 &&1917.67 &\\
16&1923.6 &&1928.33 &\\
17&1933.8 &&1937.33 &\\
18&1944.2 &&1947.42 &\\
19&1954.33	&&1958.25 &\\
20&1964.83 &&1968.92 &\\
21&1976.25	 &&1980 &\\
22&1986.75&&1989.58 &\\
23&1996.42 &&2000.33 &\\
24&2009.0 &&2013 &\\
\hline
\end{tabular}
\end{table}

Figure~B1 complements the PSD data of Figure~5 by showing a
``waterfall diagram'', or modified Gabor transform (Majkowski et al. 2014), 
of the combined maxima and minima data of 
Schove, supplemented by the two ``lost cycles''.
Besides the clear peak around 11 years, we observe also
the typical shortening of the cycles before the  
``lost cycles''.

\begin{figure}
\includegraphics[width=\linewidth]{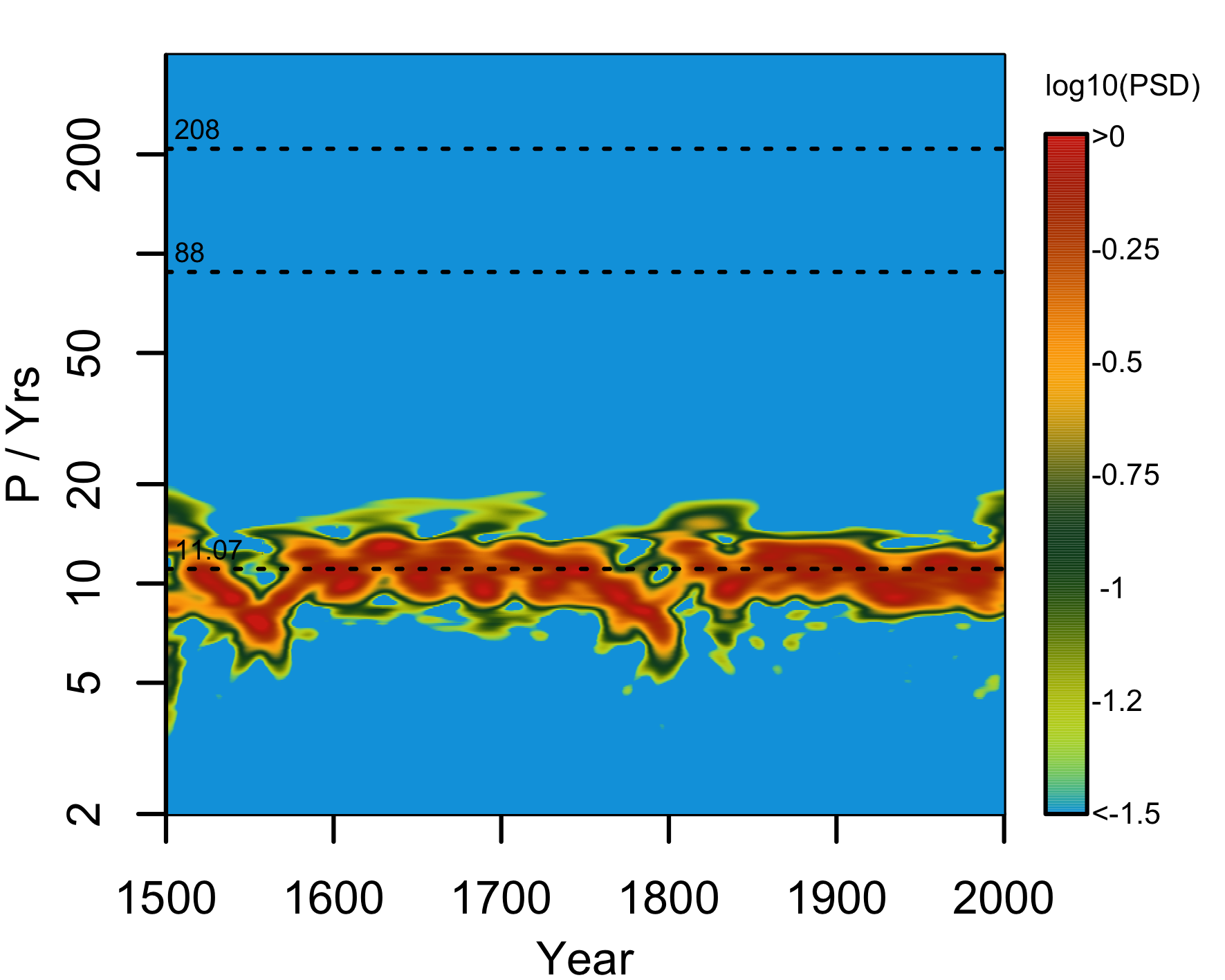}
\caption{``Waterfall diagram'' (modified Gabor transform)
of Schove's data with the two ``lost cycles'' included.}
\label{figa1}
\end{figure}

\section{Phase analysis, Lomb-Scargle diagram, and waterfall diagram 
for radioisotopes}

In Figure~C1 we show the phase diagrams for the $^{14}$C (a,b) and $^{10}$Be (c,d)
data, produced by means of the phaselet method with bandwidths
$b=$(85 yr)$^{-1}$ (a,c) and (29 yr)$^{-1}$ (b,d).  Evidently, the 
resulting band structure is as far
not as clear as in Figure~4.
Given that the Schove/Hathaway data underlying Figure~4 should be
considered as rather safe down to 1700, say, we tend to attribute the
problem to the quality of the radioisotope data.
From Figure~C1d we also learn that
for the recognition of the phase jump around 1565
an appropriately narrow observation window is recommendable,
which might have consequences for the interpretation of 
missing phase jumps in Vos et al. (2004).

\begin{figure}
\includegraphics[width=\linewidth]{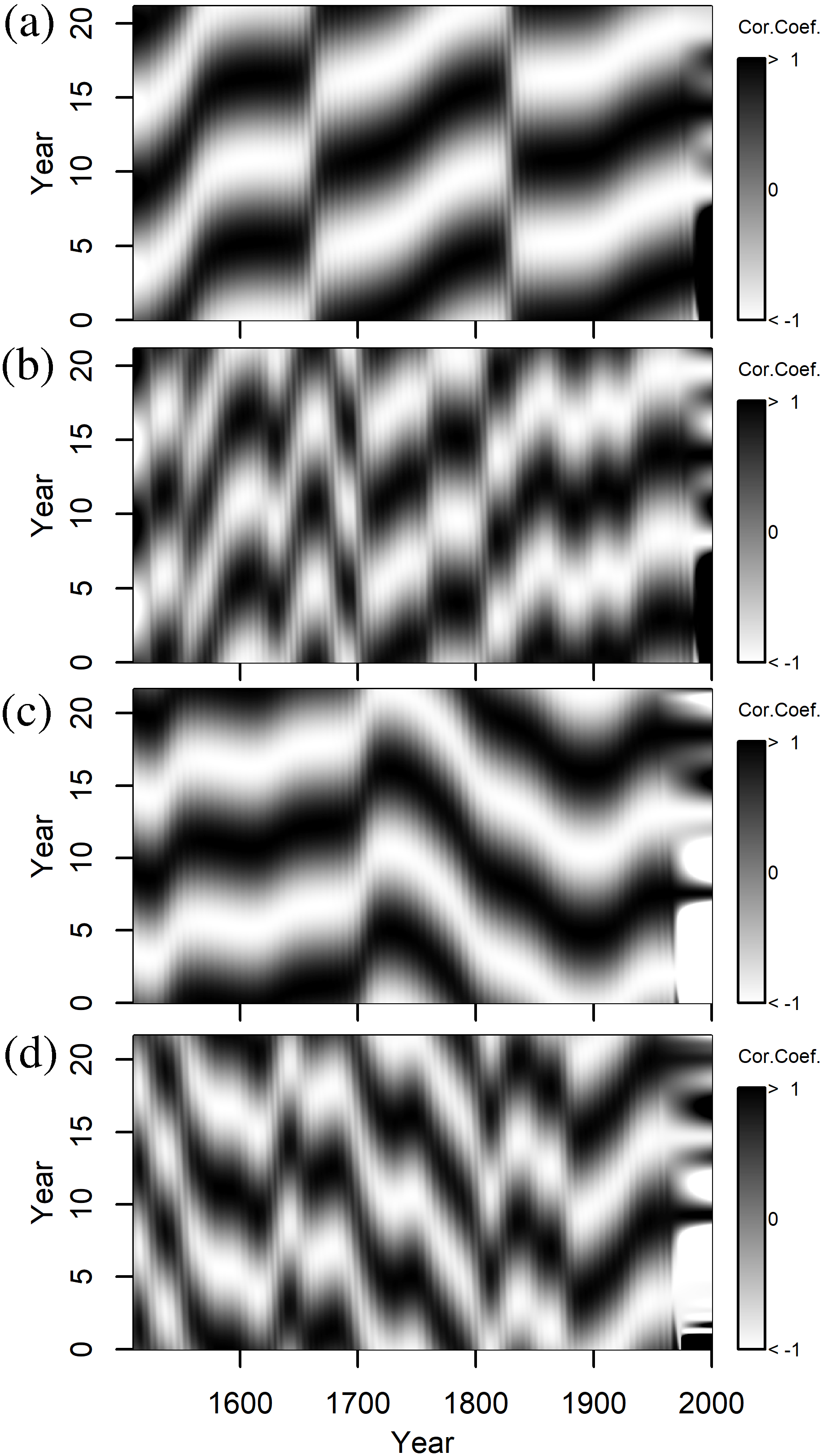}
\caption{Phase diagrams of the radioisotope date: $^{14}$C (a,b) 
and $^{10}$Be (c,d), as 
generated by means of the phaselet method with bandwidths
$b=$(85 yr)$^{-1}$ (a,c) and (29 yr)$^{-1}$ (b,d).}
\label{figb1}
\end{figure}

The corresponding Lomb-Scargle diagrams is shown in Figure~C2, with
a slightly shifted peak around 10.5 years, and another nice peak 
of the Suess-de Vries type (appr. 200 years). 
Additionally, Figure~C3  shows the
``waterfall diagram'' (modified Gabor transform) 
of the $^{14}$C data.

\begin{figure}
\includegraphics[width=\linewidth]{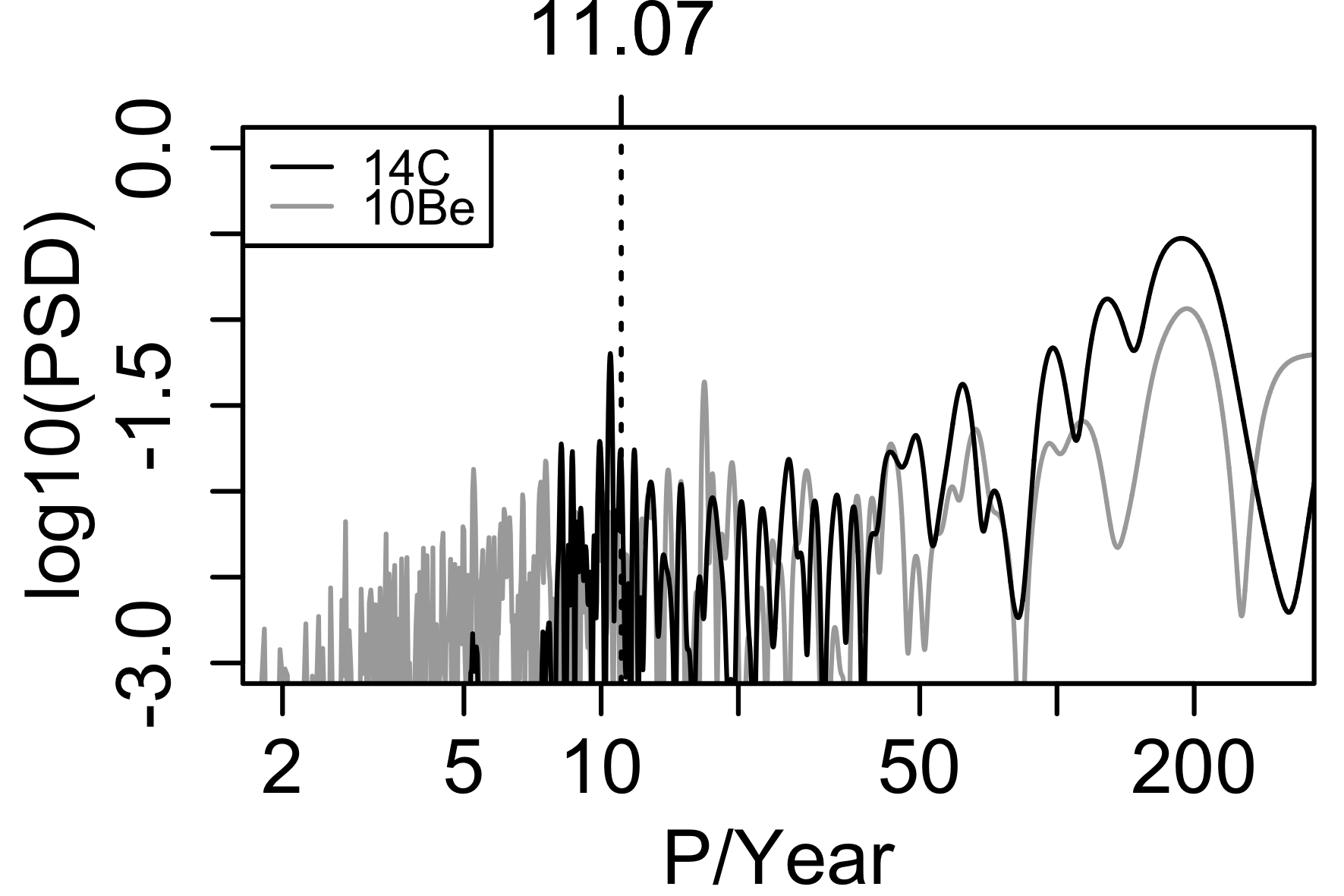}
\caption{Lomb-Scargle diagram of the radioisotope date: $^{14}$C 
and $^{10}$Be.
}
\label{figb2}
\end{figure}

\begin{figure}
\includegraphics[width=\linewidth]{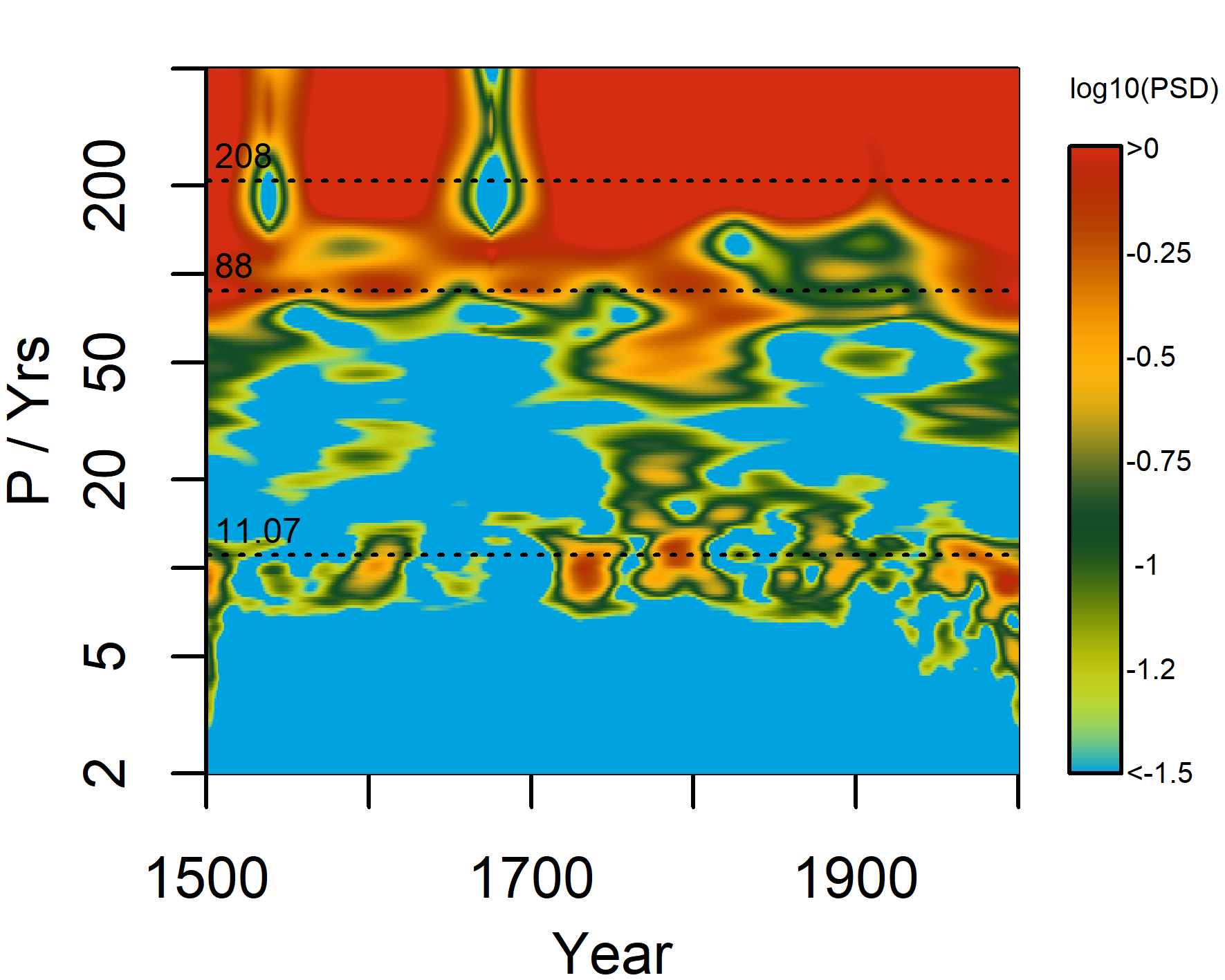}
\caption{``Waterfall diagram'' (modified Gabor transform)
of the $^{14}$C data.}
\label{figb3}
\end{figure}
\end{document}